\documentclass{article}
\usepackage[utf8]{inputenc}
\usepackage[english]{babel}
\usepackage{amsmath,amsthm,amssymb}
\usepackage{qcircuit}
\usepackage{physics}
\usepackage{caption}
\usepackage{subcaption}
\usepackage[ruled]{algorithm2e}
\usepackage{appendix}
\usepackage{floatrow}
\usepackage{arxiv}
\usepackage{graphicx}
\usepackage{tikz}

\newcommand{\B}{\mathbb{B}}
\newcommand{\N}{\mathbb{N}}

\newcommand{\BB}[1][n]{\B^{\B^{#1}}}

\newfloatcommand{capbtabbox}{table}[][\FBwidth]
\newtheorem{thrm}{Theorem}[section]

\newtheorem{prop}{Proposition}[section]
\newtheorem{definition}{Definition}[section]

\theoremstyle{definition}

\newenvironment{myproof} {\vspace{-2em}\textbf{Proof}: \newline}{\hfill$\blacksquare$}

\newcommand{\inputgroupvp}[6]{\POS"#1,1"."#2,1"."#1,1"."#2,1"!C*+<#3>\frm{\{}, \POS"#1,1"."#2,1"."#1,1"."#2,1"*!C!<#6,#4>=<0em>{#5}}

\newcommand{\filter}[1]{{_{\uparrow}#1}}

\title{Exact Learning with Tunable Quantum Neural Networks and a Quantum Example Oracle}

\author{
    Viet Pham Ngoc\\
    Imperial College London\\London, United Kingdom\\
    \texttt{viet.pham-ngoc17@imperial.ac.uk}
    \And
    Herbert Wiklicky\\
    Imperial College London\\London, United Kingdom\\
    \texttt{h.wiklicky@imperial.ac.uk}
}

\begin{document}
\maketitle

\begin{abstract}
In this paper, we study the tunable quantum neural network architecture in the quantum exact learning framework with access to a uniform quantum example oracle. We present an approach that uses amplitude amplification to correctly tune the network to the target concept. We applied our approach to the class of positive $k$-juntas and found that $O(n^22^k)$ quantum examples are sufficient with experimental results seemingly showing that a tighter upper bound is possible.
\end{abstract}


\section{Introduction}
Introduced in \cite{Angluin1988}, the model of exact learning is, together with QPAC learning, one of the main framework of learning theory. Initially, the learner is given access to two different oracles: the membership oracle and the equivalence oracle. Suppose that $c$ is the target concept. Given $x$, a call to the membership oracle will return $c(x)$ and given the learner's current hypothesis $h$, the equivalence oracle will output a randomly chosen $x \in \B^n$ such that $h(x) \neq c(x)$ should such $x$ exists The goal is then to produce a hypothesis $h^*$ such that it is always equal to $c$. Given the impracticality of the equivalence oracle it has become common practice to only work with the membership oracle \cite{Bshouty1998, Ding2019} or forgo these two oracle altogether and use another kind, generally a random example oracle. In this case, the preferred flavor is the uniformly distributed examples \cite{Sellie2009, Mossel2003, Haviv2016}. These last two research directions naturally gave rise to quantum generalisations of these oracles.

In this work we will study the performances of the tunable quantum neural network as introduced in \cite{PhamNgoc2020} in the quantum exact learning framework when the learner has access to the quantum version of the uniform example oracle. To do so we devised a training algorithm that leverages amplitude amplification and fine tuned it by learning generic Boolean functions. We show that in this case it performs better than a naive algorithm that does not use amplitude amplification. Finally we adapted this algorithm to learn the class of $k$-juntas and found that it requires less examples than what can be found in the literature.


\section{Quantum Exact Learning}
As said previously, quantum exact learning is the extension of the exact learning framework to quantum oracles. We define the quantum membership oracle \cite{Arunachalam2017, Arunachalam2021, Servedio2004} and the uniform quantum example oracle \cite{Atici2007} and formalise the training target in the quantum exact learning framework. 

\begin{definition}
    Let $n \in \N$ and suppose that the target concept is $c \in \BB$. Then the quantum membership oracle $\mathbf{MO}(c)$ will have the following action:
    \begin{equation}
        \forall x \in \B^n, \forall b \in \B, \ \mathbf{MO}(c)\ket{x,b} = \ket{x, b \oplus c(x)}
    \end{equation}
\end{definition}

On the other hand, the uniform quantum example oracle is defined as:
\begin{definition}\label{def:exact_super}
    Let $n \in \N$ and $c \in \BB$ be the target concept, then a query to the quantum example oracle $\mathbf{EX}(c)$ will return the superposition $\ket{\psi(c)}$ such that:
    \begin{equation}\label{eq:exact_super}
        \ket{\psi(c)} = \frac{1}{\sqrt{2^n}}\sum_{x \in \B^n}{\ket{x,c(x)}}
    \end{equation}
\end{definition}

By using either one of these oracles, the goal of the learner in the exact learning framework is:
\begin{definition}
    Let $n \in \N$ and $\mathcal{C} \subseteq \BB$ be a concept class. An algorithm is said to be an exact learner for $\mathcal{C}$ if for every $c \in \mathcal{C}$, it outputs a hypothesis $h \in \BB$ such that, with high probability:
    \begin{equation}
        \forall x \in \BB, h(x) = c(x)
    \end{equation}
    This notion of high probability is to be defined. While some authors use a threshold of 2/3 \cite{Arunachalam2017}, we chose to use 0.95.
\end{definition}
In this contribution, we aim at performing exact learning with a learner having access to the uniform quantum example oracle.


\section{Learning Algorithm}

\subsection{Preliminary Analysis}
Let $n \in \N$ and $c \in \BB$. As stated previously, we have access to the oracle $\mathbf{EX}(c)$, a query to which will produce the superposition $\ket{\psi(c)}$ given in Equation \ref{eq:exact_super}.

Now suppose that, in its current state (denoted $\mathbf{T}(h)$), the network is expressing $h \in \BB$ then:
\begin{equation}\label{eq:T+psi}
    \mathbf{T}(h)\ket{\psi(c)} = \frac{1}{\sqrt{2^n}}\sum_{h(x) = c(x)}{\ket{x}\ket{0}} + \frac{1}{\sqrt{2^n}}\sum_{h(x) \neq c(x)}{\ket{x}\ket{1}}
\end{equation}
So from Equation \ref{eq:T+psi}, it comes that the number of misclassified inputs and the probability of measuring the read-out qubit in state $\ket{1}$ are directly proportional:
\begin{prop}
    Let $E(h) = \{x \in \B^n \ | \ h(x) \neq c(x)\}$ be the set of misclassified inputs and $P_1$, the probability of measuring the read-out qubit in state $\ket{1}$, then:
    \begin{equation} \label{eq:P1 and E}
        P_1 = \frac{|E(h)|}{2^n}
    \end{equation}
    In particular, we have: 
    \begin{equation}
        \min_{P_1 > 0}P_1 = \frac{1}{2^n}
    \end{equation}
\end{prop}

The idea is thus to use amplitude amplification to boost this probability and hence measure as many misclassified inputs as possible before updating the network. In this setup, the diffusion operator is as follows:
\begin{definition}\label{def:diffusion_naive}
    Let $n \in \N$ and $c \in \BB$ be the target concept. Suppose that the network is expressing the function $h \in \BB$ in its current state, denoted $\mathbf{T}(h)$, we set:
    \begin{equation}
        \mathcal{X}_{\alpha_0}(c) = \mathbf{I} - 2\ket{\psi(c)}\bra{\psi(c)}
    \end{equation}
    \begin{equation}
        \mathbf{U}(h) = \mathbf{T}(h)
    \end{equation}
    And
    \begin{equation}
        \mathcal{X}_G = \mathbf{I}^{\otimes n} \otimes \mathbf{Z}
    \end{equation}
    This allows us to define the diffusion operator:
    \begin{equation}
        \mathbf{Q}(c,h) = - \mathbf{U}(h)\mathcal{X}_{\alpha_0}(c)\mathbf{U}^{\dagger}(h)\mathcal{X}_G
    \end{equation}
\end{definition}
Because the minimum non-zero probability of measuring a misclassified input is known, we can also determine the maximum number of iterations of the diffusion operator needed to appropriately amplify the amplitudes of the inputs of interest: 
\begin{definition}\label{def:theta_min_k_max}
    Let $n \in \N$, we define:
    \begin{equation}
        \theta_{\min} = \arcsin\left(\frac{1}{\sqrt{2^n}}\right)
    \end{equation}
    And:
    \begin{equation}
        m_{\max} = \arg \min_{k \in \N} \left| (2k+1)\theta_{\min} - \frac{\pi}{2}\right|
    \end{equation}
\end{definition}

We now show the following property:
\begin{prop}
    Let $P_1$ be a non-zero probability of measuring the read-out qubit in state $\ket{1}$ in this setup. Let us denote $P_1^{(m)}$ this same probability but after $m$ rounds of amplitude amplification. Then there exists $m_{1/2} \leq m_{\max}$ such that:
    \begin{equation}
        P_1^{(m_{1/2})} \geq \frac{1}{2}
    \end{equation}
\end{prop}
\begin{myproof}
    Let $\theta = \arcsin(\sqrt{P_1})$, then because $P_1$ is non-zero, we have:
    \begin{equation}
        \frac{1}{2^n} \leq P_1 \leq 1
    \end{equation}
    Thus:
    \begin{equation}
        \theta_{\min} \leq \theta \leq \frac{\pi}{2}
    \end{equation}
    If $\frac{\pi}{4} \leq \theta \leq \frac{\pi}{2}$, then we can take $m_{1/2} = 0$. Otherwise, there exist $m > 0$ such that:
    \begin{equation}
        \frac{\pi}{4(2m+1)} \leq \theta < \frac{\pi}{4(2m-1)}
    \end{equation}
    So:
    \begin{equation}
        \frac{\pi}{4} \leq (2m+1)\theta < \frac{\pi}{4} +\frac{\pi}{2(2m-1)} \leq \frac{3\pi}{4}
    \end{equation}
    Hence after $m$ rounds of amplitude amplification, we have:
    \begin{equation}
        \frac{1}{2} \leq \sin^2((2m+1)\theta) = P_1^{(m)}
    \end{equation}
    Let us denote this $m$, $m_{1/2}$, then because $P_1 \geq \frac{1}{2^n}$, we also have $m_{1/2} \leq m_{\max}$
\end{myproof}

The general idea for the algorithm is the following:
\begin{enumerate}
    \item \label{item:first} E is the set of misclassified inputs, initialised with the empty set
    \item For $m \in [0,m_{\max}]$, perform $m$ rounds of AA:
    \begin{itemize}
        \item Perform $s$ measurements
        \begin{itemize}
            \item If 1 is measured on the read-out qubit, add the measurement of the $n$ first qubits to $E$ the set of misclassified inputs
        \end{itemize}
    \end{itemize}
    \item \label{item:stop} Update the TNN according to $E$ and restart from \ref{item:first}.
    \item The algorithm stops if $E$ remains empty when \ref{item:stop}. is reached
\end{enumerate}
In order to ensure that the algorithm correctly terminates 95\% of the time, we give a condition on the number of measurements $s$.  

\begin{definition}\label{def:N_k}
    Let $m \in [0, m_{\max}]$ and $s \in \N$, we define $N_m$ the number of times the read-out qubit is measured in state $\ket{1}$ after $m$ rounds of amplification and over $s$ measurements. Then:
    \begin{equation}
        N_m \sim \text{B}\left(s, P_1^{(m)}\right)
    \end{equation}
\end{definition}

Using Definition \ref{def:N_k}, we can now estimate the probability for the algorithm to terminate incorrectly:

\begin{prop}\label{prop:min_s}
    The probability that the algorithm stops when there are still misclassified inputs is given by:
    \begin{equation}
        P(N_0 = 0, \ldots, N_{m_{\max}} = 0 \mid P_1 > 0)
    \end{equation}
    If $s \geq 5$, then
    \begin{equation}
        P(N_0 = 0, \ldots, N_{m_{\max}} = 0 \mid P_1 > 0) \leq 0.05
    \end{equation}
\end{prop}
\begin{myproof}
    We have:
    \begin{equation}
        P(N_0 = 0, \ldots, N_{m_{\max}} = 0 \mid P_1 > 0) \leq P(N_{m_{1/2}} = 0 \mid P_1 > 0)
    \end{equation}
    But 
    \begin{equation}
        P(N_{m_{1/2}} = 0 \mid P_1 > 0) = (1-P_1^{(m_{1/2})})^s \leq \frac{1}{2^s}
    \end{equation}
    So by taking $s \geq 5$, we do have:
    \begin{equation}
        P(N_0 = 0, \ldots, N_{m_{\max}} = 0 \mid P_1 > 0) \leq 0.05
    \end{equation}
\end{myproof}

Given this condition on $s$, we will now specify further the learning algorithm through the task of learning a generic Boolean function.

\subsection{Learning a Generic Boolean Function}\label{sec:generic_exact}
Let $n \in \N$, in this section, we aim at learning $c \in \BB$ without assuming any property or structure on $c$, that is we take the concept class $\mathcal{C}$ to be $\BB$. This task allows us to specify the learning algorithm to train a TNN in the exact learning framework. To do so we compare it to a naive approach. Without assuming anything about the target concept $c$, the naive way to correctly learn it, is to evaluate $c(x)$ for all $x \in \BB$. Given that we only have access to a uniform quantum example oracle and not a quantum membership oracle the expected number of queries to measure all the inputs is given by:

\begin{thrm}\label{th:coupon}
    Let $K$ be a set with $|K| = N$ and $\{k_i\}$ be uniformly sampled elements from $K$. Now let $s \in \N$ such that:
    \begin{equation}
        \{k_i\}_{1 \leq i \leq s} = K
    \end{equation}
    This problem is known as the coupon collector's problem \cite{Flajolet1992} and it can be shown that:
    \begin{equation}
        E(s) = N \ln(N)
    \end{equation}
\end{thrm}

\subsubsection{Naive algorithm}
Using Theorem \ref{th:coupon} we propose the naive learning algorithm described in Algorithm \ref{alg:naive}. With this algorithm, during each update phase, $\theta(n 2^n)$ queries to the oracle are done. Because after each update, we should have measured all of the misclassified inputs, the result from \cite{PhamNgoc2020} can be applied and the total number of updates is in $\Theta(n)$. The total number of queries to the example oracle can thus be evaluated to be in $\Theta(n^2 2^n)$. 

\begin{algorithm}[h]
    \caption{Naive algorithm}\label{alg:naive}
    \KwData{$\ket{\psi(c)}$ and $\mathbf{T}$ the network with all gates initialised to $\mathbf{I}$}
    \KwResult{Tuned network $\mathbf{T}$ expressing $c$}
    $E \leftarrow [0]$\;
    $s \leftarrow \lfloor 2^n \ln(2^n) \rfloor$\;
    \While{$E \neq \varnothing$}{
        $E \leftarrow []$\;
        \For{$1 \leq i \leq S$}{
            Measure $\mathbf{T}\ket{\psi(c)}$\;
            \If{1 is measured on the ancillary qubit}{
                Add the first $n$ qubits to $E$\;
            }
        }
        \For{$u \in E$}{
            Update $\mathbf{G}_u$ in $\mathbf{T}$\;
        }
    }
\end{algorithm}

One property of this algorithm is that it puts an emphasis on the misclassified inputs. However within these $2^n \ln(2^n)$ measurements, a mix of misclassified and correctly classified inputs will be measured. By using amplitude amplification, the results of the measurements can then be focused on the inputs of interest. The idea is thus to adapt the number of samples to the number of amplification rounds in order to measure just what is necessary.

\subsubsection{A first improvement}
Let $N_{err} \in [1, 2^n]$ be the number of misclassified inputs and $\theta_{err} \in \left]0, \frac{\pi}{2}\right]$ defined by:
\begin{equation}
    \theta_{err} = \arcsin\left(\sqrt{\frac{N_{err}}{2^n}}\right) = \arcsin\left(\sqrt{P_1}\right)
\end{equation}
Then there exists $m_{err} \in \N$ such that:
\begin{equation}
    \theta_{err} \in \left]\frac{\pi}{2(2m_{err}+3)}, \frac{\pi}{2(2m_{err}+1)}\right]
\end{equation}
This yields:
\begin{equation}
    N_{err} \in \left]\sin^2\left(\frac{\pi}{2(2m_{err}+3)}\right)2^n, \sin^2\left(\frac{\pi}{2(2m_{err}+1)}\right)2^n\right]
\end{equation}
And after $m_{err}$ rounds of amplification, the probability $P_1^{(m_{err})}$ of measuring the readout qubit in state $\ket{1}$ is now:
\begin{equation}
    P_1^{(m_{err})} \in \left]\sin^2\left(\frac{\pi}{2}-\frac{\pi}{2m_{err}+3}\right), 1\right]
\end{equation}
For $m$ sufficiently large enough, if $N_{err} \in \left]\sin^2\left(\frac{\pi}{2(2m_{err}+3)}\right)2^n, \sin^2\left(\frac{\pi}{2(2m_{err}+1)}\right)2^n\right]$, then after $m$ rounds of amplification, the results of the measurements will mostly be misclassified inputs. Applying Theorem \ref{th:coupon} and taking into account the condition on the number of samples from Property \ref{prop:min_s} we define the following:

\begin{definition}\label{def:s_k_first_attempt}
    Let $n\in \N$ and $m_{\max}$ as in Definition \ref{def:theta_min_k_max}. Then for $m \in [0 \ldots m_{\max}]$, we denote:
    \begin{equation}
        N_m = \sin^2\left(\frac{\pi}{2(2m+3)}\right)2^n
    \end{equation}
    And
    \begin{equation}
        s_m = \max(5, N_m \ln(N_m))
    \end{equation}
\end{definition}
The total number of samples during an update phase is thus: $\sum_{m=0}^{m_{\max}}{s_m}$ which is to be compared to $2^n \ln(2^n)$

\begin{prop}
    Let $n \in \N$, $m_{\max}$ as in Definition \ref{def:theta_min_k_max} and $s_m$ defined as in Definition \ref{def:s_k_first_attempt}. Then for $n$ large enough:
    \begin{equation}
        \sum_{k=0}^{m_{\max}}{s_m} < 2^n\ln(2^n)
    \end{equation}
\end{prop}
\begin{myproof}
    For the sake of clarity, for $m \in [0 \ldots m_{\max}]$, we define:
    \begin{equation}
        u_m = \sin^2\left(\frac{\pi}{2(2m+3)}\right)
    \end{equation}
    So that $N_m = u_m2^n$. Instead of directly showing $\sum_{m=0}^{m_{\max}}{s_m} < 2^n\ln(2^n)$, we will instead show:
    \begin{equation}
        \sum_{m=0}^{m_{\max}}{(N_m\ln(N_m)+5)} < 2^n\ln(2^n)
    \end{equation}
    For the first sum:
    \begin{equation}
        \sum_{m=0}^{m_{\max}}{N_m\ln(N_m)} = \sum_{m=0}^{m_{\max}}{\ln(N_m^{N_m})} = \ln(\prod_{m=0}^{m_{\max}}N_m^{N_m})
    \end{equation}
    And
    \begin{align}
        \prod_{m=0}^{m_{\max}}N_m^{N_m} &= \prod_{m=0}^{m_{\max}}(u_m2^m)^{N_m} = \prod_{m=0}^{m_{\max}}{u_m^{N_m}}\prod_{m=0}^{m_{\max}}{(2^n)^{N_m}} = \prod_{m=0}^{m_{\max}}{u_m^{N_m}}\prod_{m=0}^{m_{\max}}{(2^n)^{u_m2^n}}\\
        &= \prod_{m=0}^{m_{\max}}{u_m^{N_m}}\prod_{m=0}^{m_{\max}}{((2^n)^{2^n})^{u_m}}
    \end{align}
    Because $u_m \leq 1$, we have $\prod_{m=0}^{m_{\max}}{u_m^{N_m}} \leq 1$, hence:
    \begin{equation}
        \prod_{m=0}^{m_{\max}}N_m^{N_m} \leq \prod_{m=0}^{m_{\max}}{((2^n)^{2^n})^{u_m}}
    \end{equation}
    But:
    \begin{equation}
        \prod_{m=0}^{m_{\max}}{((2^n)^{2^n})^{u_m}} = ((2^n)^{2^n})^{\sum_{m=0}^{m_{\max}}u_m}
    \end{equation}
    And:
    \begin{align}
        \sum_{m=0}^{m_{\max}}{u_m} &= \sum_{m=0}^{m_{\max}}{\sin^2\left(\frac{\pi}{2(2m+3)}\right)}\\
        &\leq \sum_{m=0}^{m_{\max}}{\left(\frac{\pi}{2(2m+3)}\right)^2} \\
        &\leq \frac{\pi^2}{4}\sum_{m=1}^{m_{\max}+1}{\frac{1}{(2m+1)^2}} \\
        &\leq \frac{\pi^2}{4} \left(\sum_{m=1}^{+\infty}{\frac{1}{m^2}}-\sum_{m=1}^{+\infty}{\frac{1}{(2m)^2}}-1\right)\\
        &\leq \frac{\pi^2}{4} \left(\frac{\pi^2}{6}-\frac{\pi^2}{24}-1\right)\\
        &\leq 0.58
    \end{align}
    This means:
    \begin{equation}
        \prod_{m=0}^{m_{\max}}N_m^{N_m} \leq ((2^n)^{2^n})^{0.58}
    \end{equation}
    Hence:
    \begin{equation}
        \sum_{m=0}^{m_{\max}}{N_m\ln(N_m)} \leq 0.58 \times 2^n\ln(2^n)
    \end{equation}
    Now for the second sum:
    \begin{equation}
        \sum_{m=0}^{m_{\max}}{5} = 5(m_{\max}+1)
    \end{equation}
    But $m_{\max} \approx \sqrt{2^n}$ so for $n$ large enough:
    \begin{equation}
        \sum_{m=0}^{k_{\max}}{5} \approx 5 \sqrt{2^n} \lll 2^n\ln(2^n)
    \end{equation}
    Its contribution is thus negligible in the overall sum and we have:
    \begin{equation}
        \sum_{m=0}^{m_{\max}}{s_m} < \sum_{m=0}^{m_{\max}}{(N_m\ln(N_m)+5)} < 2^n\ln(2^n) 
    \end{equation}
\end{myproof}

\begin{figure}[h]
    \centering
    \begin{subfigure}{0.495\textwidth}
        \centering
        \includegraphics[scale=0.5]{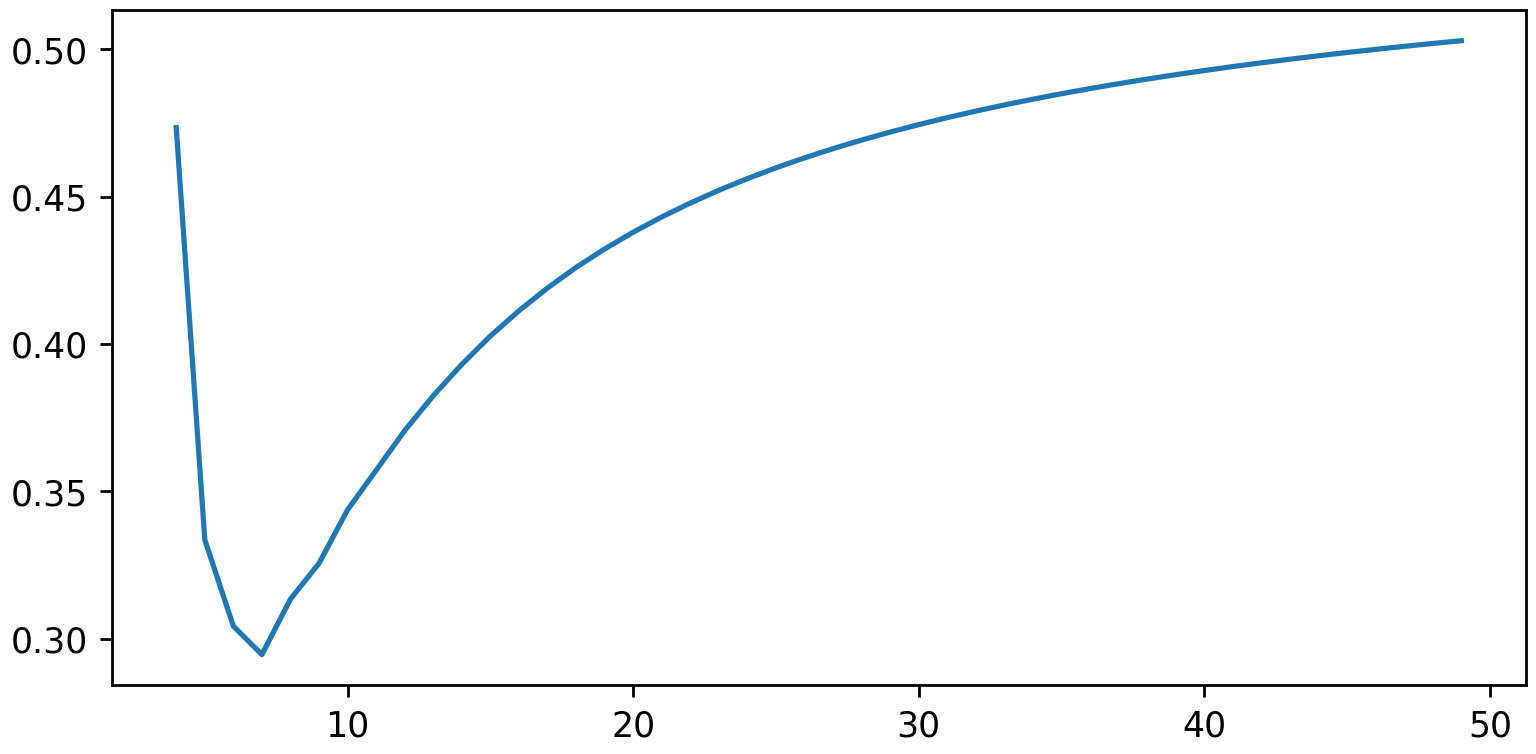}
        \caption{$k$ incremented by 1}
        \label{fig:ratio_0_1}
    \end{subfigure}
    \hfill
    \begin{subfigure}{0.495\textwidth}
        \centering
        \includegraphics[scale=0.5]{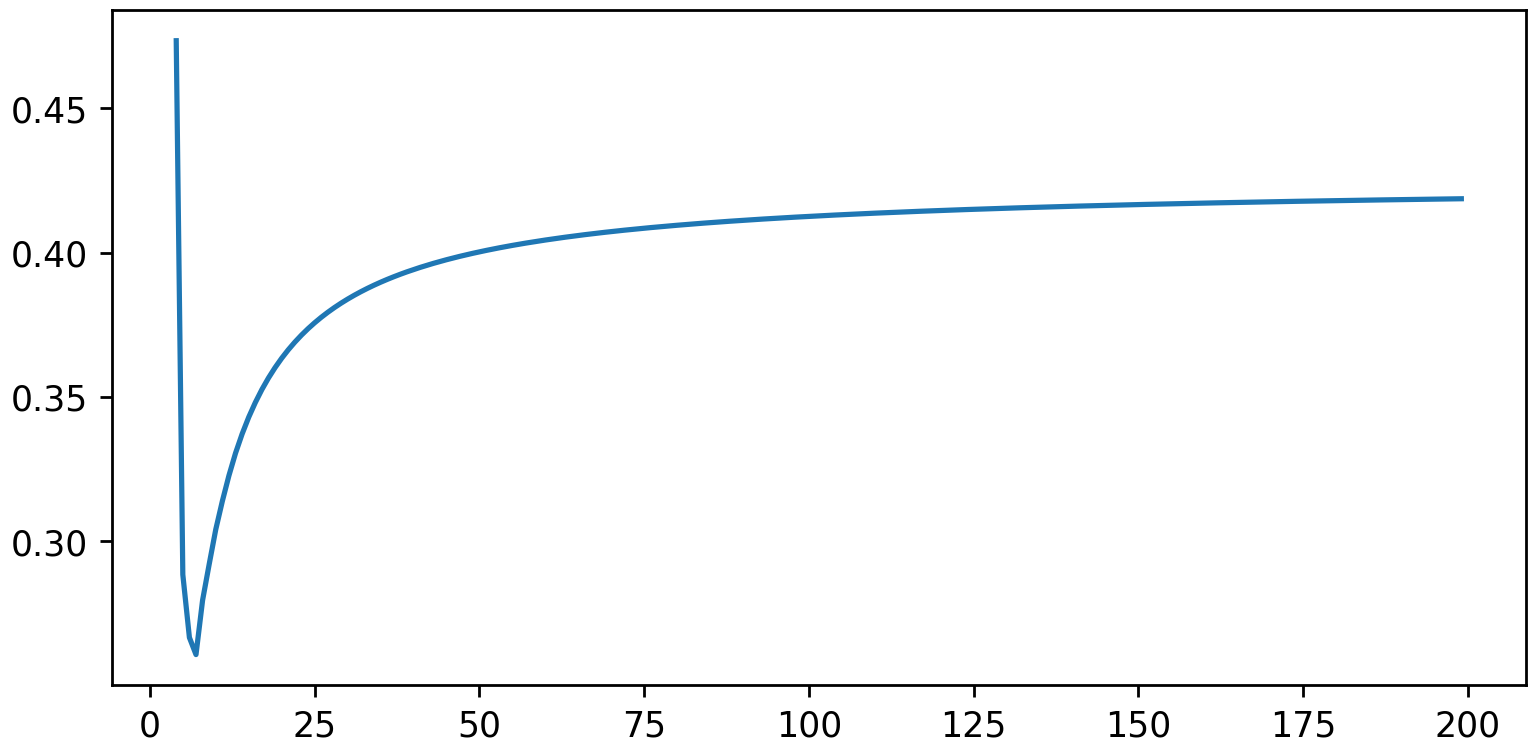}
        \caption{$k$ incremented with powers of 2}
        \label{fig:ratio_0_2}
    \end{subfigure}
    
    \caption{Ratio $\sum_{m=0}^{m_{\max}}{s_m}/2^n\ln(2^n)$ as a function of $n$ for $n \geq 4$. In Figure \ref{fig:ratio_0_1} $m$ has been incremented by 1 while in Figure \ref{fig:ratio_0_2}, $m$ has been incremented with powers of 2}
    \label{fig:ratio_0}
\end{figure}

\begin{figure}[h]
    \centering
    $$
        \Qcircuit @R=1em @C=1em @!R{
            & {/^n} \qw    & \multigate{1}{\mathbf{T}(h)} & \qw    & \multigate{1}{\mathbf{Q}^m(c,h)}  & \qw & \meter\\
            & \qw          & \ghost{\mathbf{T}(h)}        & \qw   & \ghost{\mathbf{Q}^m(c,h)}         & \qw & \meter
            \inputgroupvp{1}{2}{1em}{1.2em}{\ket{\psi(c)}}{2.2em}
        }
    $$
    \caption{Quantum circuit for the improved algorithm}
    \label{fig:circuit_improved}
\end{figure}
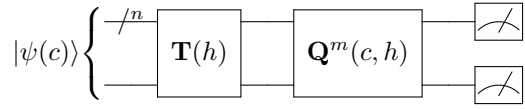

In order to have a better idea of the reduction in oracle queries introduced by this algorithm, we have plotted in Figure \ref{fig:ratio_0}  the ratio, as a function of $n$ for $n \geq 4$, between the total number of samples and $2^n \ln(2^n)$.
As shown in Figure \ref{fig:ratio_0_1}, for $n$ large enough, the total number of samples is approximately halved when compared to the naive algorithm. To reduce further this number and speed up the training algorithm, we decided to increment $m$, the number of amplification rounds, not by 1 but with powers of 2. The ratio of the total number of queries using this scheme, compared to the naive algorithm is given in Figure \ref{fig:ratio_0_2}. In this case, we can see that it has been divided by more than 2. For these reasons, the increment schedule for $m$ will now be:
\begin{equation}
    [0, 1, 2, 4, \ldots, m_{\max}]
\end{equation}
The improved procedure is detailed in Algorithm \ref{alg:improve1}.
\begin{algorithm}
    \caption{Improved algorithm}\label{alg:improve1}
    \KwData{$\ket{\psi(c)}$ and $\mathbf{T}$ the network with all gates initialised to $\mathbf{I}$}
    \KwResult{Tuned network $\mathbf{T}$ expressing $c$}
    $E \gets [0]$\;
    $schedule \gets [0,1,2,4,\ldots,m_{\max}]$\;
    \While{$E \neq \varnothing$}{
        $E \leftarrow []$\;
        \For{$m \in schedule$}{
            $N \leftarrow \sin^2\left(\frac{\pi}{2(2m+3)}\right)2^n$\;
            $S \leftarrow \max(5, N \ln(N))$\;
            Perform $m$ rounds of amplification\;
            \For{$1 \leq i \leq S$}{
                Measure \;
                \If{1 is measured on the ancillary qubit}{
                    Add the first $n$ qubits to $E$\;
                }
            }
        }
        \For{$u \in E$}{
            Update $\mathbf{G}_u$ in $\mathbf{T}$\;
        }
    }
\end{algorithm}
If $m$ is the number of amplification rounds, the circuit used in Algorithm \ref{alg:improve1} is depicted in Figure \ref{fig:circuit_improved}.

\subsubsection{A further refinement}
One issue with this algorithm resides in the fact that for $n \in \N$ and $0 \leq m \leq m_{\max}$, the range of misclassified inputs to be covered is:
\begin{equation}\label{eq:range_first_improve}
    \left]\sin^2\left(\frac{\pi}{2(2m+3)}\right)2^n, \sin^2\left(\frac{\pi}{2(2m+1)}\right)2^n\right]
\end{equation}
So as $m$ increases, the range will decrease. This means that approximating the number of misclassified inputs with the lower bound becomes more accurate as $m$ increases. However, for small $m$, this approximation is not as accurate. For example, for $m=0$, the range given in Equation \ref{eq:range_first_improve} becomes:
\begin{equation}
    \left]\frac{2^n}{4}, 2^n\right]
\end{equation}
This range can be refined by adding a second ancillary qubit, initialised in state $\ket{0}$ and a rotation gate acting on this qubit and controlled by the first ancillary qubit. We give a more accurate definition of this gate:
\begin{definition}
    Let $m_0 \in \N$, we define: 
    \begin{equation}
        \theta_{m_0} = \frac{\pi}{2(2m_0+1)}
    \end{equation}
    And we denote $\mathbf{CR}_{m_0}$ the controlled rotation (around the $y$-axis) gate such that:
    \begin{equation}
        \mathbf{CR}_{m_0}\ket{10} = \cos\left(\theta_{m_0}\right)\ket{10} + \sin\left(\theta_{m_0}\right)\ket{11}
    \end{equation}
\end{definition}
Then if $c \in \BB$ is the target concept and the network is expressing $h\in \BB$ in its current state $\mathbf{T}(h)$, we have:
\begin{equation}\label{eq:CR_k0T}
    \mathbf{CR}_{m_0}\mathbf{T}(h)\ket{\psi(c)}\ket{0} = \sin\left(\theta_{m_0}\right) \frac{1}{\sqrt{2^n}}\sum_{h(x) \neq c(x)}{\ket{x}\ket{11}} + \ket{\perp}
\end{equation}
So the inputs of interest are now marked with the two ancillary qubits being in $\ket{11}$ and the probability $P_{11}$ of measuring such inputs is now such that:
\begin{equation}\label{eq:interval_P11}
    P_{11} \in \left[0, \sin^2\left(\theta_{m_0}\right)\right]
\end{equation}
And in this case:
\begin{equation}
    \min_{P_{11}>0} = \sin^2\left(\theta_{m_0}\right)\frac{1}{2^n}
\end{equation}
We adapt Definitions \ref{def:diffusion_naive} and \ref{def:theta_min_k_max} to take into account this modification:
\begin{definition}\label{def:diffusion_k_0}
    Let $n \in \N$, $c \in \BB$ be the target concept and $m_0 \in \N$. Suppose that the network is expressing the function $h \in \BB$ in its current state $\mathbf{T}(h)$. We denote:
    \begin{equation}
        \mathcal{X}_{\alpha_0}(c) = (\mathbf{I}-2\ket{\psi(c)}\bra{\psi(c)})\otimes \mathbf{I}
    \end{equation}
    \begin{equation}
        \mathbf{U}_{m_0}(h) = \mathbf{CR}_{m_0}\mathbf{T}(h)
    \end{equation}
    And
    \begin{equation}
        \mathcal{X}_G = \mathbf{I}^{\otimes n+1}\otimes \mathbf{Z}
    \end{equation}
    The diffusion operator $\mathbf{Q}_{m_0}(c,h)$ is then defined as:
    \begin{equation}
        \mathbf{Q}_{m_0}(c,h) = -\mathbf{U}_{m_0}(h)\mathcal{X}_{\alpha_0}(c)\mathbf{U}^{\dagger}_{m_0}(h)\mathcal{X}_G
    \end{equation}
\end{definition}
We now define the quantities that are related to this modified process.
\begin{definition}\label{def:theta_min_k_max_refined}
    Let $n \in \N$ and $m_0 \in \N$, we denote:
    \begin{equation}
        \theta_{\min,m_0} = \arcsin\left(\sin\left(\theta_{m_0}\right)\frac{1}{\sqrt{2^n}}\right)
    \end{equation}
    And:
    \begin{equation}
        m_{\max,m_0} = \arg \min_{m \in \N} \left|(2m+1)\theta_{\min,m_0}-\frac{\pi}{2}\right|
    \end{equation}
\end{definition}
Let $m_0 \in \N$. Now let $\theta_{err} \in \left]0, \frac{\pi}{2}\right]$ such that:
\begin{equation}
    \theta_{err} = \arcsin(\sqrt{P_{11}})
\end{equation}
Then there exists $m \in \N$ such that:
\begin{equation}
    \theta_{err} \in \left]\frac{\pi}{2(2m+3)}, \frac{\pi}{2(2m+1)}\right]
\end{equation}
But according to Equation \ref{eq:interval_P11}, we have $P_{11} \leq \sin^2\left(\theta_{m_0}\right)$ so it comes that:
\begin{equation}
    m \geq m_0
\end{equation}
Now if $N_{err}$ is the number of misclassified inputs, then Equation \ref{eq:CR_k0T} yields:
\begin{equation}
    N_{err} = \sin^2(\theta_{err})\frac{2^n}{\sin^2(\theta_{m_0})}
\end{equation}
Hence:
\begin{equation}\label{eq:interval_Nerr}
    N_{err} \in \left]\sin^2\left(\frac{\pi}{2(2m+3)}\right)\frac{2^n}{\sin^2(\theta_{m_0})}, \sin^2\left(\frac{\pi}{2(2m+1)}\right)\frac{2^n}{\sin^2(\theta_{m_0})}\right]
\end{equation}
So for $m \geq m_0$, the ratio between the two bounds of this interval is:
\begin{equation}\label{eq:upper_bound_ratio}
    1 \leq \frac{\sin^2\left(\frac{\pi}{2(2m+1)}\right)}{\sin^2\left(\frac{\pi}{2(2m+3)}\right)} \leq \frac{\sin^2\left(\frac{\pi}{2(2m_0+1)}\right)}{\sin^2\left(\frac{\pi}{2(2m_0+3)}\right)} \approx \left(\frac{m_0+3}{m_0+1}\right)^2 = \left(1+\frac{2}{m_0+1}\right)^2
\end{equation}
Where the approximation is valid for $m_0 > 0$. So as $m_0$ grows, it is possible for the bounds of the interval given in Equation \ref{eq:interval_Nerr} to become quite close. Moreover, after $m$ rounds of amplitude amplification, we have:
\begin{equation}
    P_{11}^{(m)} \in \left]\sin^2\left(\frac{2m+1}{2m+3}\frac{\pi}{2}\right), 1\right]
\end{equation}
With:
\begin{equation}
    \sin^2\left(\frac{2m+1}{2m+3}\frac{\pi}{2}\right) = \sin^2\left(\frac{\pi}{2}-\frac{\pi}{2m+3}\right) = \cos^2\left(\frac{\pi}{2m+3}\right) = 1 - \sin^2\left(\frac{\pi}{2m+3}\right) 
\end{equation}
As $m \geq m_0$, we have:
\begin{equation}
    \sin^2\left(\frac{2m+1}{2m+3}\frac{\pi}{2}\right) \geq 1 - \sin^2\left(\frac{\pi}{2m_0+3}\right)
\end{equation}
Here again, for $m_0$ large enough, it is possible for the lower bound to be very close to 1. This means that for $m_0$ large enough, whatever the number of misclassified inputs (or equivalently the probability $P_{11}$), as long as it is non-zero, it is possible to find a $0 \leq m \leq m_{\max, m_0}$ such that after $m$ rounds of amplitude amplification, the measurement process will be a random process where the probability of measuring a  misclassified input is close to 1. In addition, the number of such inputs can more accurately be approximated by the lower bound of the interval given in Equation \ref{eq:interval_Nerr}. However, it does not suffice to take $m_0$ very large. Indeed, as $m_{\max, m_0}$ increases with $m_0$, so does the total number of samples. A good choice for this parameter is thus one that allows for the behaviour previously described while ensuring that the number of samples remains small enough. 

\begin{figure}[h]
    \centering
    \includegraphics{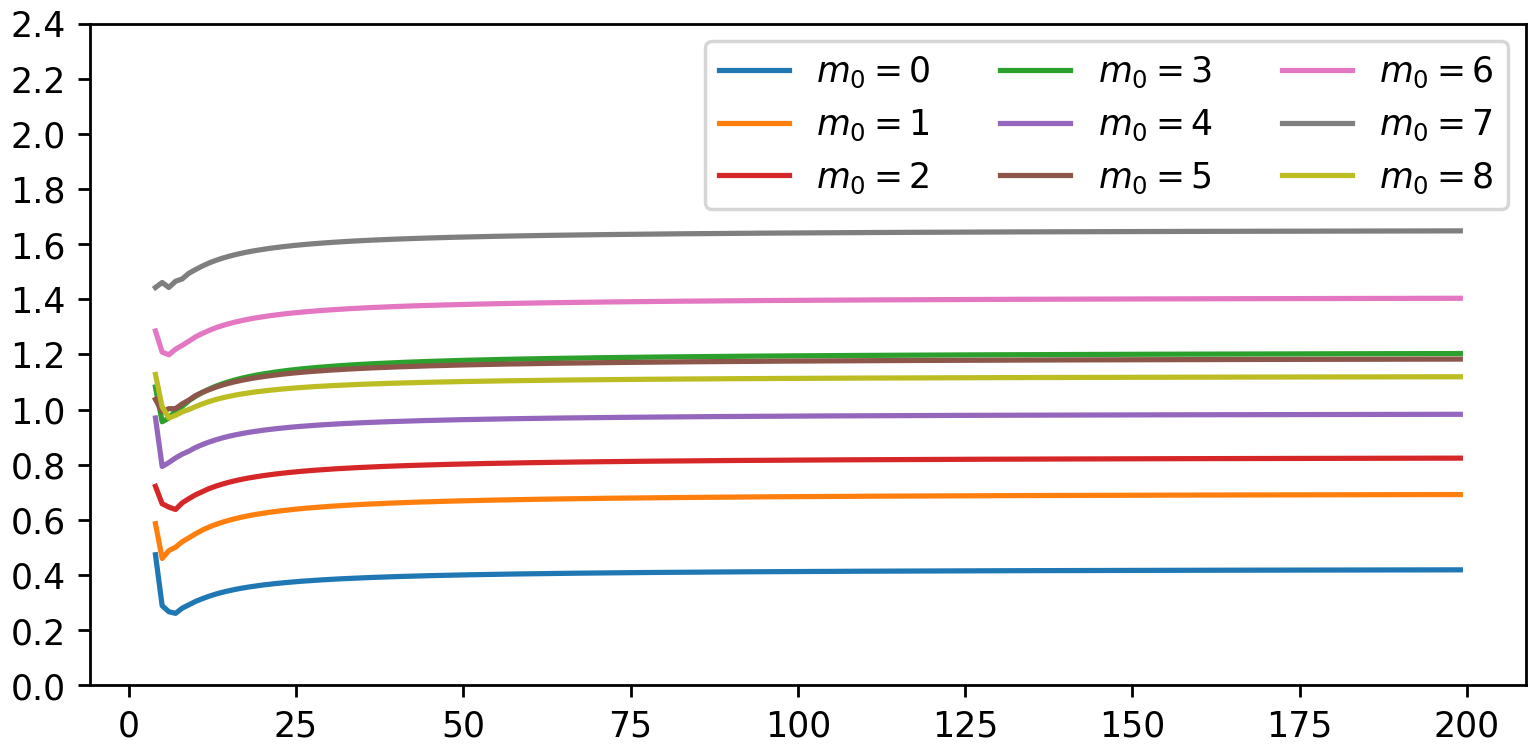}
    \caption{Ratio $S_{m_0}/2^n\ln(2^n)$ as a function of $n$ for different values of $m_0$}
    \label{fig:samples_k_0}
\end{figure}

\begin{definition}\label{def:samples_refined}
    Let $n \in \N$, $m_0 \in \N$ and $m_{\max,m_0}$ as in Definition \ref{def:theta_min_k_max_refined}. For $m_0 \leq m \leq m_{\max, k_0}$ we define:
    \begin{equation}
        N_{m,m_0} = \sin^2\left(\frac{\pi}{2(2m+3)}\right)\frac{2^n}{\sin^2(\theta_{m_0})}
    \end{equation}
    And
    \begin{equation}
        s_{m,m_0} = \max\left(N_{m,m_0} \ln(N_{m,m_0}),5\right)
    \end{equation}
\end{definition}
The total number of samples during an update phase is thus $\sum_{m=m_0}^{m_{\max}}{s_{m,m_0}}$ but in order to speed up the algorithm we chose to keep the schedule introduced in the first improvement hence:
\begin{definition}
    Let $n \in \N$, $m_0 \in \N$, $m_{\max,m_0}$ as in Definition \ref{def:theta_min_k_max_refined} and $s_{m,m_0}$ as in Definition \ref{def:samples_refined}, then the total number of samples $S_{m_0}$during an update phase is given by:
    \begin{equation}
        S_{m_0} = s_{m_0,m_0}+\sum_{m_0 < 2^p < m_{\max,m_0}}{s_{m,m_0}} + s_{m_{\max,m_0},m_0}
    \end{equation}
\end{definition}
To guide the choice for a suitable $m_0$, the ratio $S_{m_0}/2^n\ln(2^n)$ has been plotted for different values of $n \in \N$ and $_m0 \in \N$ in Figure \ref{fig:samples_k_0}.
From this figure, it is apparent that the values $0 \leq m_0 \leq 4$ are suitable with $m_0 = 0$ corresponding to the first improvement introduced earlier. Algorithm \ref{alg:refined} describes the training algorithm that will be used to train the network. 
\begin{algorithm}
    \caption{Refined algorithm}\label{alg:refined}
    \KwData{$m_0$, $\ket{\psi(c)}$ and $\mathbf{T}$ the network with all gates initialised to $\mathbf{I}$}
    \KwResult{Tuned network $\mathbf{T}$ expressing $c$}
    $E \leftarrow [0]$\;
    $schedule \leftarrow [m_0, 2^{\lfloor\log_2{m_0}\rfloor+1}, 2^{\lfloor\log_2{m_0}\rfloor+2}, \ldots,m_{\max,m_0}]$\;
    \While{$E \neq \varnothing$}{
        $E \leftarrow []$\;
        \For{$m \in schedule$}{
            $N \leftarrow \frac{\sin^2\left(\frac{\pi}{2(2k+3)}\right)}{\sin^2\left(\frac{\pi}{2(2k_0+1)}\right)}2^n$\;
            $S \leftarrow \max(5, N \ln(N))$\;
            Perform $m$ rounds of amplification\;
            \For{$1 \leq i \leq S$}{
                Measure\;
                \If{1 is measured on the first ancillary qubit}{
                    Add the first $n$ qubits to $E$\;
                }
            }
        }
        \For{$u \in E$}{
            Update $\mathbf{G}_u$ in $\mathbf{T}$\;
        }
    }
\end{algorithm}
Notice that when measuring, we still look at the first ancillary qubit. Indeed, because of the controlled rotation gate $\mathbf{CR}_{m_0}$, a misclassified input will result in the ancillary qubits being measured either in state $\ket{10}$ or $\ket{11}$. While the latter is used in the amplification process to refine it, the former still corresponds to a state of interest and it would be a waste to ignore it during the measurements.

\begin{figure}
    \centering
    $$
        \Qcircuit @R=1em @C=1em @!R{
            & {/^n} \qw    & \multigate{1}{\mathbf{T}(h)} & \qw   & \qw                & \multigate{2}{\mathbf{Q}_{m_0}^m(c,h)}  & \qw & \meter\\
            & \qw          & \ghost{\mathbf{T}(h)}        & \qw   & \ctrl{1}           & \ghost{\mathbf{Q}_{m_0}^m(c,h)}         & \qw & \meter\\
             \lstick{\ket{0}}   & \qw          & \qw                           & \qw   & \gate{\mathbf{R}_{m_0}} & \ghost{\mathbf{Q}_{m_0}^m(c,h)}         & \qw & \meter
             \inputgroupvp{1}{2}{1em}{1.2em}{\ket{\psi(c)}}{2.2em}
        }
    $$
    \caption{Quantum circuit for the improved algorithm}
    \label{fig:circuit_refined}
\end{figure}
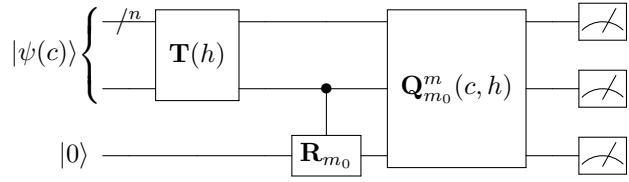

For $m_0 \in \N$ and $m \in \N$, the circuit being measured is depicted in Figure \ref{fig:circuit_refined}.

\subsection{Implementation}
\begin{figure}[h]
    \centering
    \begin{subfigure}[t]{0.49\textwidth}
        \includegraphics[width=\textwidth]{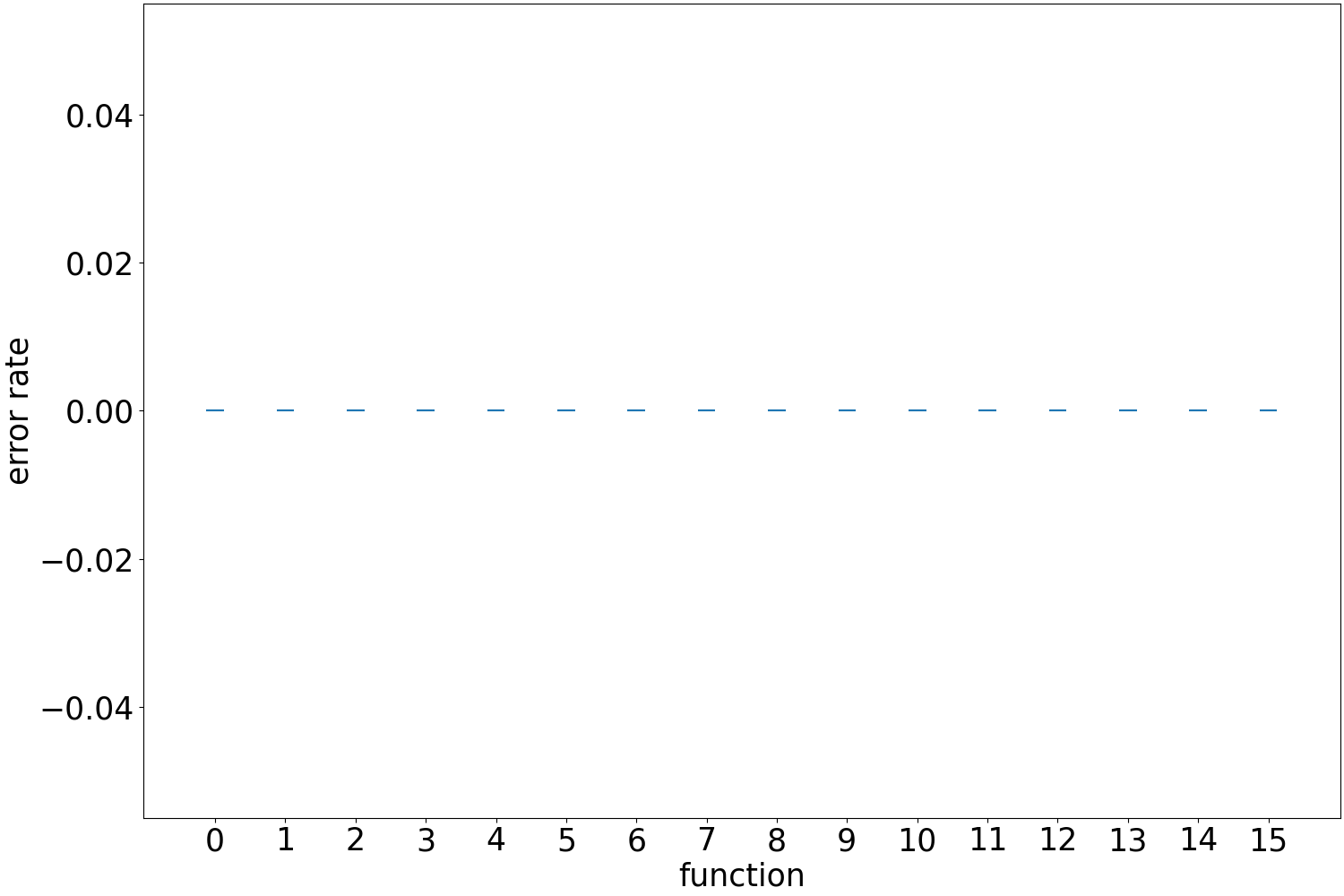}
        \caption{Algorithm using amplitude amplification for $4 \leq n \leq 8$ and $0 \leq m_0 \leq 4$}
        \label{fig:final_error_refined}
    \end{subfigure}
    \begin{subfigure}[t]{0.49\textwidth}
        \includegraphics[width=\textwidth]{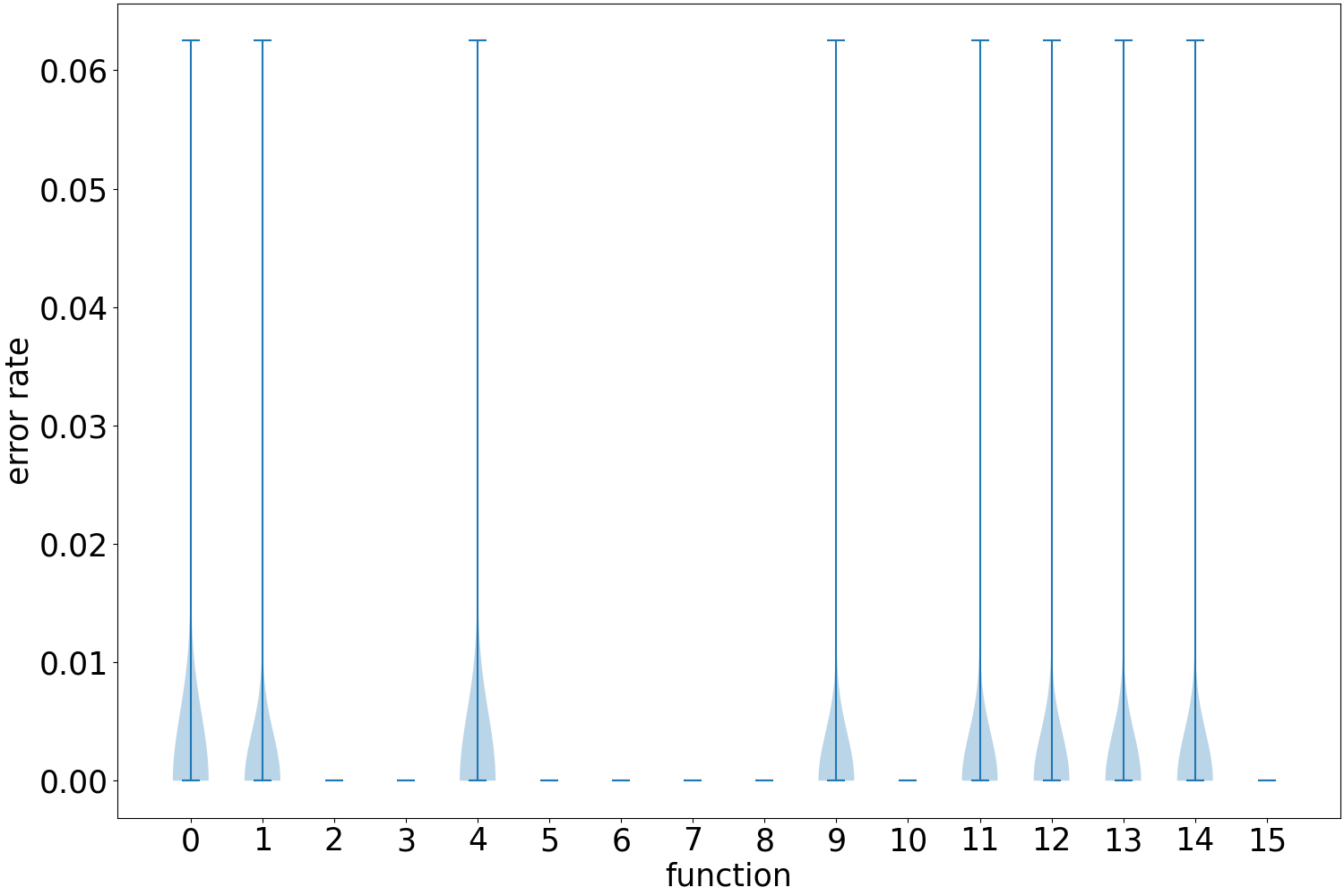}
        \caption{Naive algorithm for $n=4$}
        \label{fig:final_error_naive_4}
    \end{subfigure}
    \begin{subfigure}[t]{0.49\textwidth}
        \includegraphics[width=\textwidth]{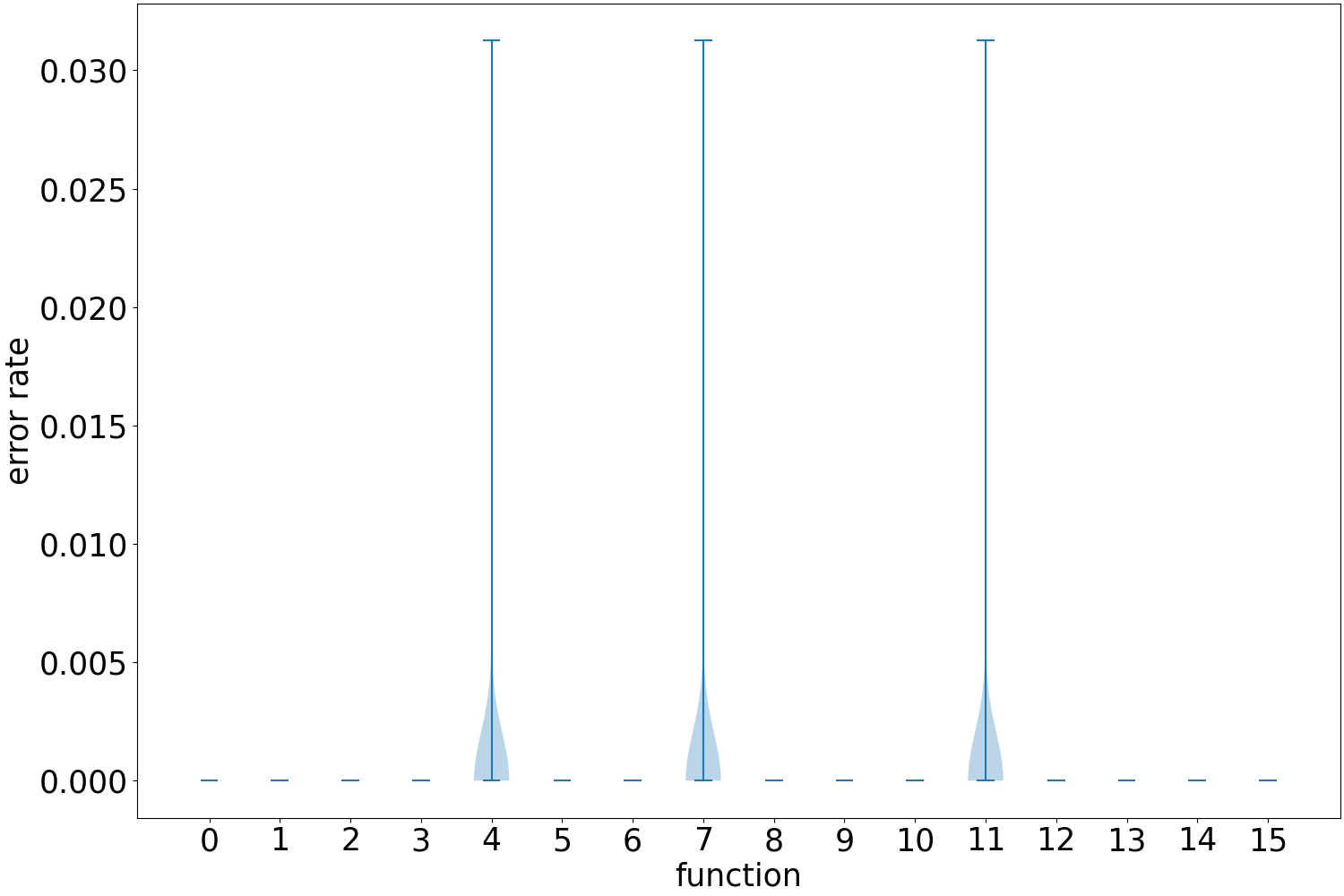}
        \caption{Naive algorithm for $n=5$}
        \label{fig:final_error_naive_5}
    \end{subfigure}
    \begin{subfigure}[t]{0.49\textwidth}
        \includegraphics[width=\textwidth]{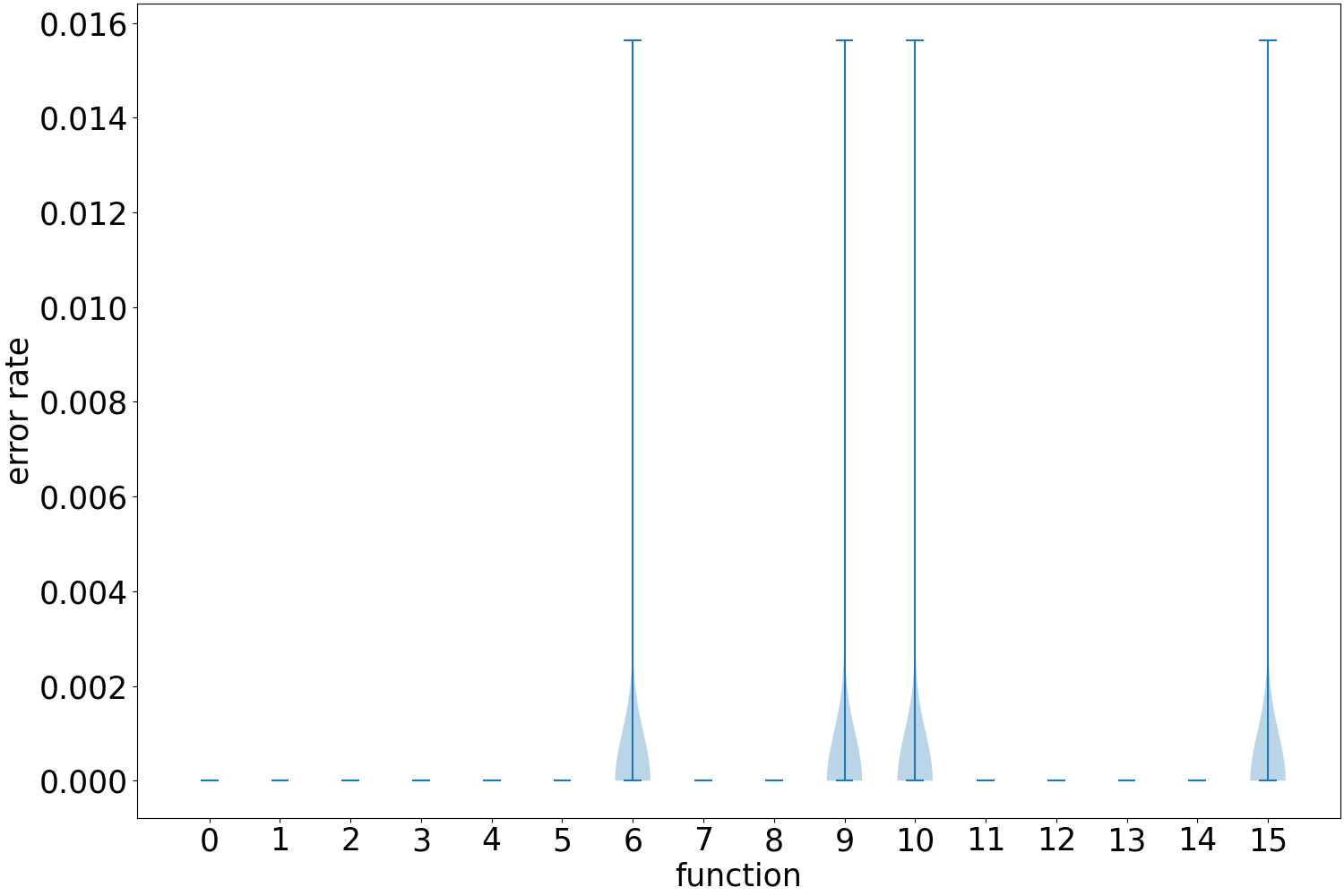}
        \caption{Naive algorithm for $n=6$}
        \label{fig:final_error_naive_6}
    \end{subfigure}
    
    \caption{Final error rate for different experiments}
    \label{fig:final_error_exact}
\end{figure}

Let $n \in \N$. In order to specify the choice of $m_0$ we have implemented Algorithm \ref{alg:refined} for $0 \leq m_0 \leq 4$ with the target being a generic Boolean function $c \in \BB$. In detail, for $4 \leq n \leq 8$, 16 target functions have been chosen randomly and for each target function, the network has been trained 50 times. Each time, the number of samples needed until the algorithm stops has been recorded as well as the final error rate. As a comparison, the naive algorithm has also been implemented and studied under the same regime. Concerning the error rate, when using amplitude amplification, whatever the values for $n$ and $m_0$, the final error rate was consistently equal to 0, as shown in Figure \ref{fig:final_error_refined}, indicating that the target function has indeed been exactly learnt. On the other hand, when training with the naive algorithm, some of the experiments for $n=4$ to $6$ failed to exactly learn the target function as depicted in Figures \ref{fig:final_error_naive_4} to \ref{fig:final_error_naive_6}.

\begin{figure}[h]
    \centering
    \includegraphics[width=0.8\textwidth]{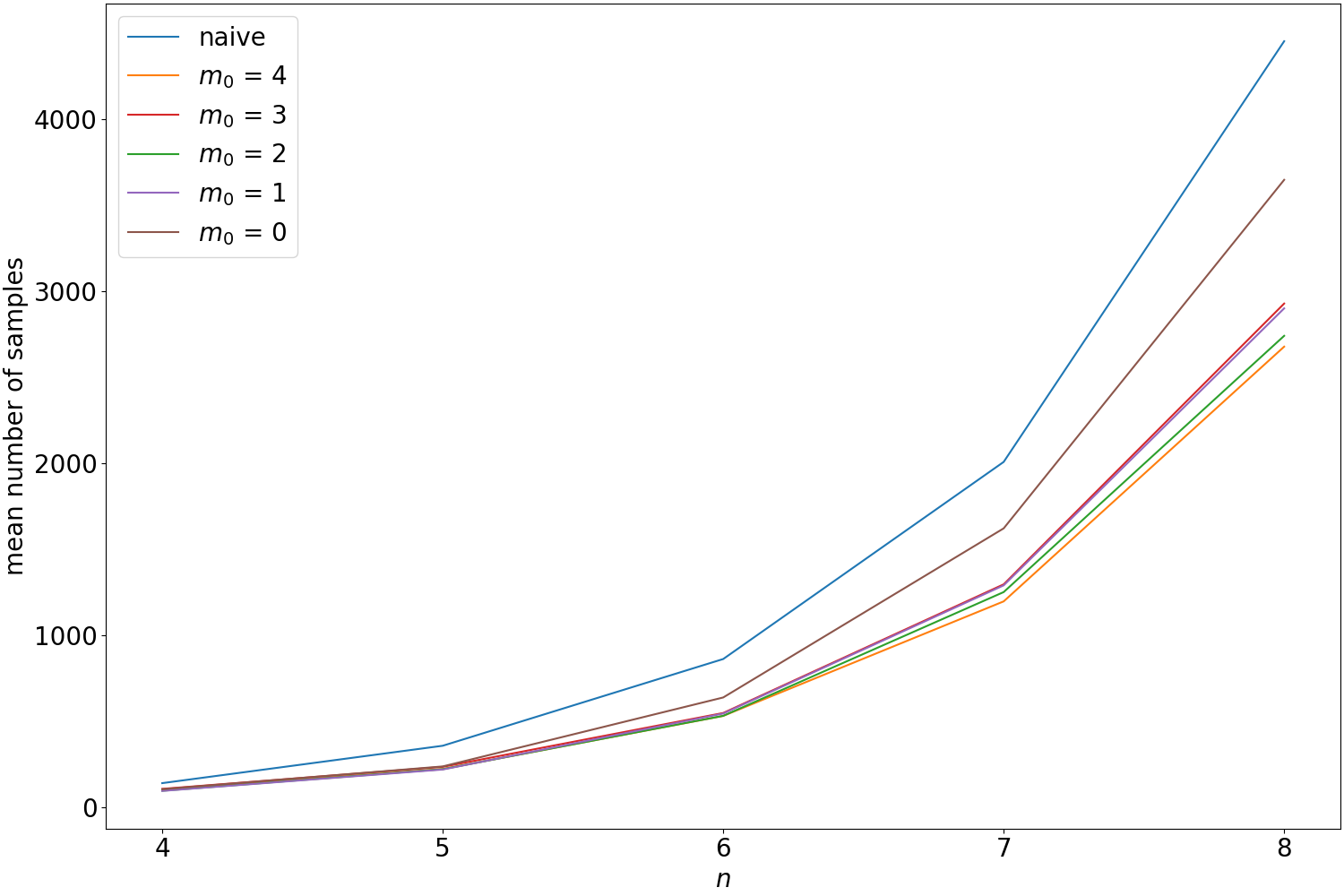}
    \caption{Mean number of samples taken over all the experiments (functions and runs) against the dimension $n$ for different values of $m_0$ ($0 \leq m_0 \leq 4$) and the naive algorithm}
    \label{fig:all_samples_exact}
\end{figure}

Hence, when only looking at the goal of exactly learning, the algorithm using amplitude amplification performs better than the naive algorithm and thus for the different values of $m_0$ that have been selected. To further confirm the advantage of the refined algorithm over the naive one, we put our focus on the number of samples required to learn the target functions. For each chosen $m_0$, the mean number of samples taken over all the experiments, i.e. all the 16 functions and the 50 runs by function, has been plotted against the dimension of the input space $n$ for $4 \leq n \leq 8$ in Figure \ref{fig:all_samples_exact}. As a point of comparison, the same metric has been plotted for the naive algorithm.

What transpires from this comparison is that for the selected $m_0$, the number of necessary samples is considerably lower than for the naive algorithm, with the gap increasing as the input dimension increases. Two values for $m_0$ particularly stand out: $m_0=2$ and $m_0=4$ as they lead to the lowest number of samples with the plot for $m_0=4$ being below the one for $m_0=2$. While this trend seems to be contradictory with Figure \ref{fig:samples_k_0}, this can be explained by the fact that the number of updates when $m_0=4$ is lower than when $m_0=2$. Seeing that the difference in terms of sample number between these two choices is relatively small, it seems that $m_0=2$ offers the best compromise between minimising the number of samples and minimising the running time of the training algorithm. In Section \ref{sec:k-juntas}, a modified version of Algorithm \ref{alg:refined} will be used with $m_0=2$ to learn $k$-juntas. The modifications will concern the number of queries during an update stage as well as the update strategy with regard to the measurement outcomes.

\section{Learning Positive $k$-Juntas}\label{sec:k-juntas}
\subsection{Description of the Concept Class and the Update Algorithm}
\begin{definition}
    Let $n \in \N^*$ and $k < n$. A Boolean function $c \in \BB$ is said to be a $k$-junta if its output only depends on at most $k$ of its input variables, these variables are then called the relevant variables. Let $\rho_c \subset [0 \ldots n-1]$ be the set of the relevant variables of $c$, then:
    \begin{equation}
        |\rho_c| \leq k
    \end{equation}
    Additionally, we define a positive $k$-junta, $c \in \BB$, as being a $k$-junta such that:
    \begin{equation}
        c(0) = 0
    \end{equation}
    And we denote $\mathcal{J}_k^+$ the class of the positive $k$-juntas. 
\end{definition}
The fact that a function is a positive $k$-junta will impose a restriction on its algebraic normal form:
\begin{prop}
    Let $n \in \N^*$ and $c \in \BB$ a positive $k$-junta. Then there exist $u_1^{(c)}, \ldots, u_p^{(c)} \in \B^n \setminus \{0\}$ with $p < 2^k$ such that:
    \begin{equation}
        \forall i \in [1,p], 1_{u_i^{c}} \subseteq \rho_c
    \end{equation}
    And:
    \begin{equation}
        c = \bigoplus_{i=1}^p{m_{u_i^{c}}}
    \end{equation}
    We will call $u_1^{(c)}, \ldots, u_p^{(c)}$ the principals of $c$.
\end{prop}
\begin{myproof}
    Let $n \in \N^*$, $k < n$ and $c \in \mathcal{J}_k^+$. For $x \in \B^n$, we denote $x_{|\rho_c}$ the restriction of $x$ to the relevant variables of $c$. As $c$ is a $k$-junta, there exist $
    c' \in \BB[|\rho_c|]$ such that:
    \begin{equation}
        \forall x \in \B^n, c(x) = c'(x_{|\rho_c})
    \end{equation}
    As $c' \in \BB[|\rho_c|]$, it has an ANF, i.e. there exist $u_1^{(c')}, \ldots, u_p^{(c')} \in \B^{|\rho_c|}$ such that:
    \begin{equation}
        c' = \bigoplus_{i=1}^p{m_{u_i^{c'}}}
    \end{equation}
    Because $c(0) = c'(0) = 0$, we also have:
    \begin{equation}
        u_1^{(c')}, \ldots, u_p^{(c')} \in \B^{|\rho_c|} \setminus \{0\}
    \end{equation}
    Now to extend $c'$ to $c$, for $i \in [1 \ldots p]$ we simply extend $u_i^{c'} \in \B^{|\rho_c|}$ to $u_i^{c} \in \B^n$ by padding with zeros where necessary. This ensures that $1_{u_i^{c}} \subseteq \rho_c$, $u_1^{(c)}, \ldots, u_p^{(c)} \in \B^n \setminus \{0\}$ and:
    \begin{equation}
        c = \bigoplus_{i=1}^p{m_{u_i^{c}}}
    \end{equation}
\end{myproof}

To understand the effect of the principals of a $k$-junta, we will introduce the notion of filter, derived from set theory:
\begin{definition}
    Let $n \in \N$ and $u \in \B^n$. We denote by $\filter{u}$ the filter generated by $u$ defined by:
    \begin{equation}
        \filter{u} = \{v \in \B^n \mid 1_u \subseteq 1_v\}
    \end{equation}
\end{definition}
By using the filters generated by the principals of a positive $k$-junta $c$, a characterisation of the inputs $v \in \B^n$ such that $c(v)=1$ can be done. 
\begin{prop}\label{prop:c(v)=1}
    Let $n \in \N^*$, and $c \in \mathcal{J}_k^{+}$. Let $u_1^{(c)}, \ldots, u_p^{(c)}$ be the principals of $c$, then for $v \in \B^n$:\\
    $c(v) = 1$ if and only if there exists an odd number of principals, $u_i^{(c)}$, such that $v$ is an element of $\filter{u_i^{(c)}}$.
\end{prop}
\begin{myproof}
    This result stems from the ANF of the function. 
\end{myproof}

In order to show the effects of updating the network, we introduce the following:
\begin{prop}
    Let $n \in \N^*$ and $c \in \mathcal{J}_k^{+}$. Let $v \in \BB$ such that $c(v) = 1$, then:
    \begin{equation}
        \forall w \in \filter{v}, c(w) \oplus m_v(w) = \overline{c(w)}
    \end{equation}
    And:
    \begin{equation}
        \forall w \notin \filter{v}, c(w) \oplus m_v(w) = c(w)
    \end{equation}
\end{prop}
\begin{myproof}
    This is a consequence of Property \ref{prop:c(v)=1}. 
\end{myproof}

Let $v \in \B^n$ such that $c(v) = 1$. Then according to Property \ref{prop:c(v)=1}:
\begin{equation}
    v \in \bigcup_{i=1}^{p}{\filter{u_i^{(c)}}}
\end{equation}
And because $\filter{v} \subset \bigcup_{i=1}^{p}{\filter{u_i^{(c)}}}$, it comes that any misclassified input is an element of the filter generated by at least one principal of the target function. By updating the network with gates controlled by misclassified inputs we thus ensure that at any time during the training process, the network will express a hypothesis $h \in \BB$ such that:
\begin{equation}
    h = \bigoplus_{u \in G}{m_u} \text{ where } G \subseteq \bigcup_{i=1}^{p}{\filter{u_i^{(c)}}}
\end{equation}
So the goal of the tuning algorithm is to gradually descend to the principals of the target function by adding to the network gates controlled by inputs of progressively lower Hamming weight. Once a gate controlled by $v$ is added, all the gates controlled by inputs in $\filter{v}$ can be trimmed from the network. 

To facilitate this process, it would be advantageous to measure misclassified inputs that are close, in terms of Hamming weight, to the principals of the target function. Indeed, by doing so, the network could be updated with gates that are closer to the principals, thus cutting down the number of update steps. This can be achieved by using, once more, amplitude amplification. The target function being a $k$-junta, we know that its principals have a Hamming weight of at most $k$. So by using AA to focus on the inputs with that property, this descent process can be facilitated. Formally, let us define the following:

\begin{definition}\label{def:diffusion<k}
    Let $n \in \N^*$ and $k < n$. We define $\mathcal{X}_{\leq k} \in U(2^{n})$ such that:
    \begin{equation}
        \forall x \in \B^n, \mathcal{X}_{\leq k}\ket{x} = 
        \begin{cases}
            - \ket{x} \text{ if } w_H(x) \leq k \\
            \phantom{-} \ket{x} \text{ else}
        \end{cases}
    \end{equation}
    Additionally, we define $\mathcal{X}_{\psi} \in U(2^{n+1})$ by:
    \begin{equation}
        \mathcal{X}_{\psi} = 2\ketbra{\psi(c)}{\psi(c)} - \mathbf{I}
    \end{equation}
    The diffusion operator for the inputs with Hamming weight of at most $k$ is then given by:
    \begin{equation}
        \mathcal{X}_{\psi}(\mathcal{X}_{\leq k} \otimes \mathbf{I})
    \end{equation}
\end{definition}
Because we know exactly how many of these inputs are in the superposition, we can determine the number of iterations of this diffusion operator are needed to appropriately amplify these inputs:
\newline

\begin{definition}\label{def:p<k}
    Let $n \in \N^*$ and $k < n$. Let $N_{\leq k}$ be the number of inputs with Hamming weight of at most $k$, then:
    \begin{equation}
        N_{\leq k} = \sum_{j=0}^{k}{\binom{n}{j}}
    \end{equation}
    
    This allows us to define $p_{\leq k}$, the number of iterations for the diffusion operator defined in Definition \ref{def:diffusion<k}:
    \begin{equation}
        p_{\leq k} = \arg\min_{p \in \N}{\left|(2p+1)\arcsin\left(\sqrt{\frac{N_{\leq k}}{2^n}}\right)-\frac{\pi}{2}\right|}
    \end{equation}
\end{definition}
We can now abstract the amplification process into a unitary:
\begin{definition}\label{def:amplification<k}
    Let $n \in \N^*$ and $k < n$. Let $\mathcal{X}_{\leq k}$ and $\mathcal{X}_{\psi}$ be the gates defined in Definition \ref{def:diffusion<k} and $p_{\leq k}$ as in Definition \ref{def:p<k}, then the operator $\mathbf{A}_{\leq k} \in U(2^{n+1})$ defined by:
    \begin{equation}
        \mathbf{A}_{\leq k} = \left[\mathcal{X}_{\psi}(\mathcal{X}_{\leq k} \otimes \mathbf{I})\right]^{p_{\leq k}}
    \end{equation}
    Will appropriately amplify the amplitudes of the inputs with Hamming weight of at most $k$.
\end{definition}

Before continuing, we must briefly redefine the diffusion operator introduced in Definition \ref{def:diffusion_k_0} for it to be compatible with this additional procedure. 

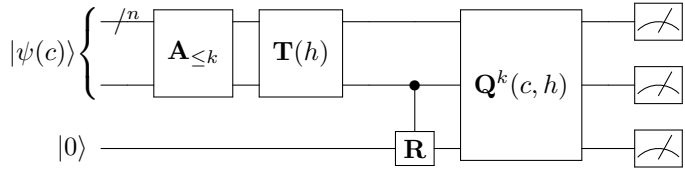
\begin{figure}[h]
    \centering
    $$
        \Qcircuit @R=1em @C=1em @!R{
            & {/^n} \qw    & \multigate{1}{\mathbf{A}_{\leq k}} & \multigate{1}{\mathbf{T}(h)} & \qw   & \qw                & \multigate{2}{\mathbf{Q}^k(c,h)}  & \qw & \meter\\
            & \qw          & \ghost{\mathbf{A}_{\leq k}}   & \ghost{\mathbf{T}(h)}     & \qw   & \ctrl{1}           & \ghost{\mathbf{Q}^k(c,h)}         & \qw & \meter\\
             \lstick{\ket{0}}   & \qw       & \qw   & \qw                           & \qw   & \gate{\mathbf{R}} & \ghost{\mathbf{Q}^k(c,h)}         & \qw & \meter
             \inputgroupvp{1}{2}{1em}{1.2em}{\ket{\psi(c)}}{2.2em}
        }
    $$
    \caption{Quantum circuit to learn $k$-juntas}
    \label{fig:circuit_k-juntas}
\end{figure}

\begin{definition}\label{def:new_diffusion}
    Let $n \in \N^*$ and $k < n$ and $\mathbf{A}_{\leq k}$ as defined in Definition \ref{def:amplification<k}. We denote:
    \begin{equation}
        \ket{\psi_{\leq k}(c)} = \mathbf{A}_{\leq k}\ket{\psi(c)}
    \end{equation}
    From this, we redefine $\mathcal{X}_{\alpha_0}$ as introduced in Definition \ref{def:diffusion_k_0} by:
    \begin{equation}
        \mathcal{X}_{\alpha_0} = (\mathbf{I}-2\ketbra{\psi_{\leq k}(c)}{\psi_{\leq k}(c)})\otimes \mathbf{I}
    \end{equation}
    The other components remaining the same (with $m_0=2$) we then have: 
    \begin{equation}
        \mathbf{Q} = - \mathbf{U}(h)\mathcal{X}_{\alpha_0}(c)\mathbf{U}(h)^{\dagger}\mathbf{X}_G
    \end{equation}
    Where $\mathbf{U}(h) = \mathbf{CR}_2\mathbf{T}(h)$.
\end{definition}
The circuit used in the learning process is depicted in Figure \ref{fig:circuit_k-juntas}.

An update phase will thus unfold as follows (to facilitate the explanation, a gate and its controlling input will be conflated). To take advantage of the filter structure, the measurements are separated into correctly classified and misclassified inputs. For each group, the inputs are sorted by increasing Hamming weight and treated in this order. For each of the misclassified inputs, the number of gates to be added, of which the misclassified input is an element of the filter, is counted. If this number is even, this input is added to the list of gates to be updated and all the gates that are in the current network and are also in this input's filter are to be removed. This update process is also performed with the correctly classified inputs but in this case, the number has to be odd in order to add the input to the list of gates to be updated. The update process described above is detailed in Algorithm \ref{alg:upd_juntas}.
\begin{algorithm}
    \caption{Update algorithm for $k$-juntas}\label{alg:upd_juntas}
    \KwData{$measurements$, the results of the measurements and $actives$, the list of the network's active gates }
    \KwResult{$to\_update$ the list of the network's gates to be updated}
    $to\_update \gets []$\;
    $errors \gets$ a list of the measured misclassified inputs where $errors[l]$ is itself the list of the inputs with Hamming weight $l$\;
    $corrects \gets$ the same as $errors$ but with the correctly classified inputs\;
    \For{$0\leq l \leq n$}{
        \For{$error \in errors[l]$}{
            $count \gets 0$\;
            \For{$upd \in to\_update$}{
                \If{$error \in \filter{upd}$}{
                    $count \gets count + 1$\;
                }
            }
            \If{$count$ is even}{
                Add $error$ to $to\_update$\;
                Add the gates in $actives$ that are also in $\filter{error}$\;
            }
        }
        \For{$correct \in corrects[l]$}{
            $count \gets 0$\;
            \For{$upd \in to\_update$}{
                \If{$correct \in \filter{upd}$}{
                    $count \gets count + 1$\;
                }
            }
            \If{$count$ is odd}{
                Add $correct$ to $to\_update$\;
                Add the gates in $actives$ that are also in $\filter{correct}$\;
            }
        }
    }
\end{algorithm}

One quantity that remains to be determined is the number of queries to the quantum example oracle during such an update phase. While it would have been possible to train the network using the sampling schedule introduced in Section \ref{sec:generic_exact}, this approach is best used when no characteristic is known of the target concept. In the case of $k$-juntas, it seems natural for the query complexity to be a function of $k$. As a result, when learning $k$-juntas, we chose to perform $2^k$ measurements after each amplification phase leading to:
\begin{prop}\label{prop:complexity_update_exact_AA}
    Let $n \in \N^*$ and $k < n$. Suppose that the concept class is $\mathcal{J}_k^+$. If the number of queries after each amplification phase is $2^k$, then during an update phase, the number of queries to the quantum example oracle is in:
    \begin{equation}
        \Theta(n2^k)
    \end{equation}
\end{prop}

This leads to the sample complexity of the whole algorithm. Thanks to Property \ref{prop:complexity_update_exact_AA} we have already established that one update phase will perform $\Theta(n2^k)$ calls to the example oracle. To determine the number of updates, we make the following observation. Let $u_i^{(c)}$ be one of the principals of the target function $c \in \mathcal{J}_k^+$. Suppose further that we update the network with a gate controlled by $v \in \filter{u_i^{(c)}}$. We are assured that during a further update step, we will correct an input $w \in \filter{u_i^{(c)}}$ such that $v \in \filter{w}$. Either because $w$ was measured randomly or because all of the elements of $\filter{v}$ have been corrected beforehand, thus ensuring that $w$ will be measured. However, thanks to the pre-amplification that focuses on the inputs with Hamming weight of at most $k$, the first event is most likely to happen.
Once the gate controlled by $w$ has been updated, the same process will repeat until we reach $u_i^{(c)}$. All in all, we have that the number of updates is in $O(n)$, and hence the query complexity is in:
\begin{equation}
    O(n^2 2^k)
\end{equation}

\subsection{Experimental Results}
\begin{figure}[h]
    \centering
    \includegraphics[width=0.5\textwidth]{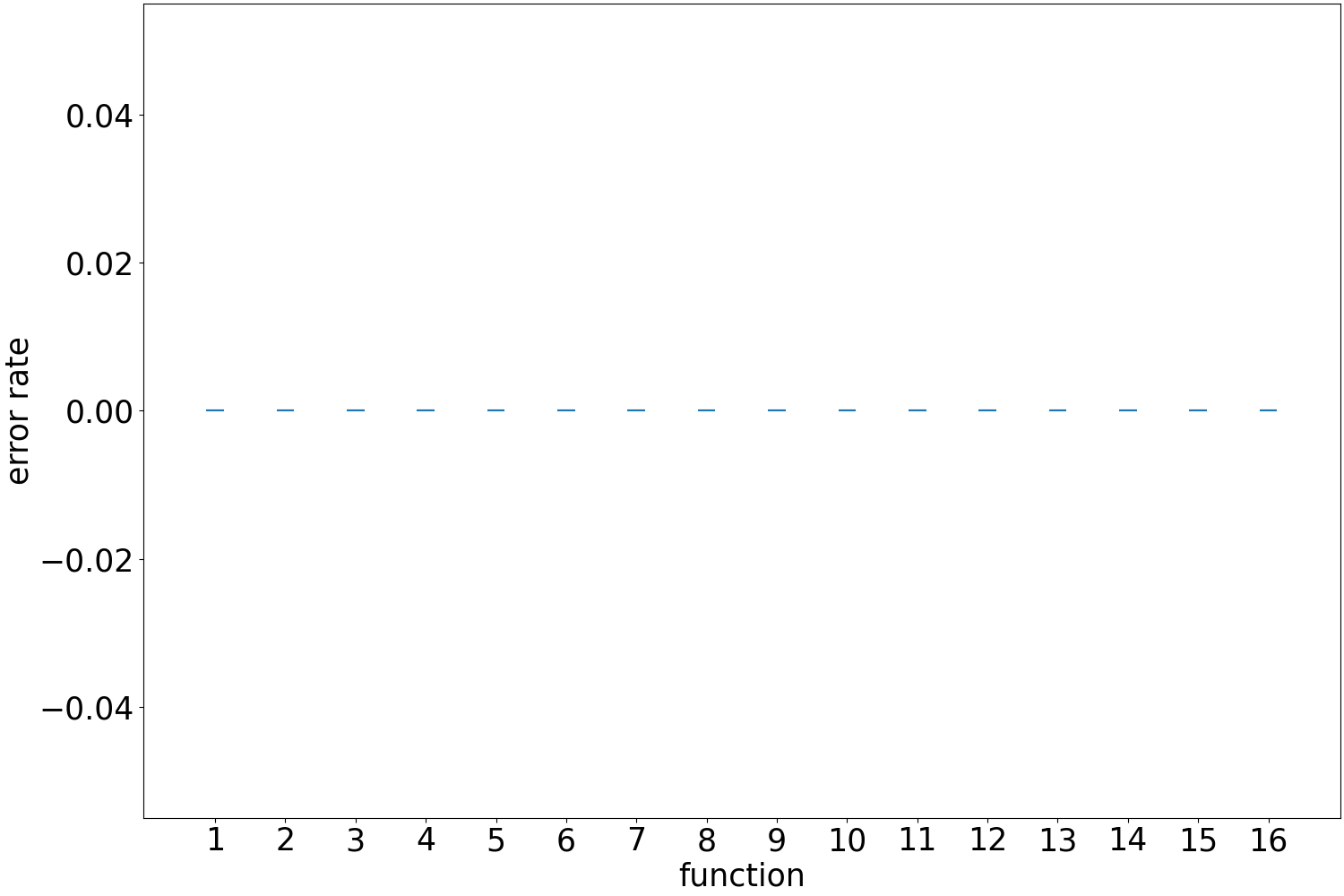}
    \caption{Final error rate for all the experiments for each pair $(n,k)$}
    \label{fig:errors_juntas}
\end{figure}

We have implemented\footnote{The code can be found here: https://github.com/vietphamngoc/exact\_AA} this learning algorithm for $n \in [5 \ldots 8]$ and $k \in [2 \ldots n-1]$. For each pair of $n$ and $k$, the network has been trained to learn 16 randomly created $k$-junta. Each training has been repeated 25 times. The final error rates for these experiments were all similar to what is reported in Figure \ref{fig:errors_juntas}.  
This figure has been plotted using violin plots, hence it shows that all of the training successfully stopped with a correctly tuned network. We now focus on the number of update steps required to reach these results. 

\begin{figure}[h]
    \centering
    \begin{subfigure}{0.49\textwidth}
        \includegraphics[width=\textwidth]{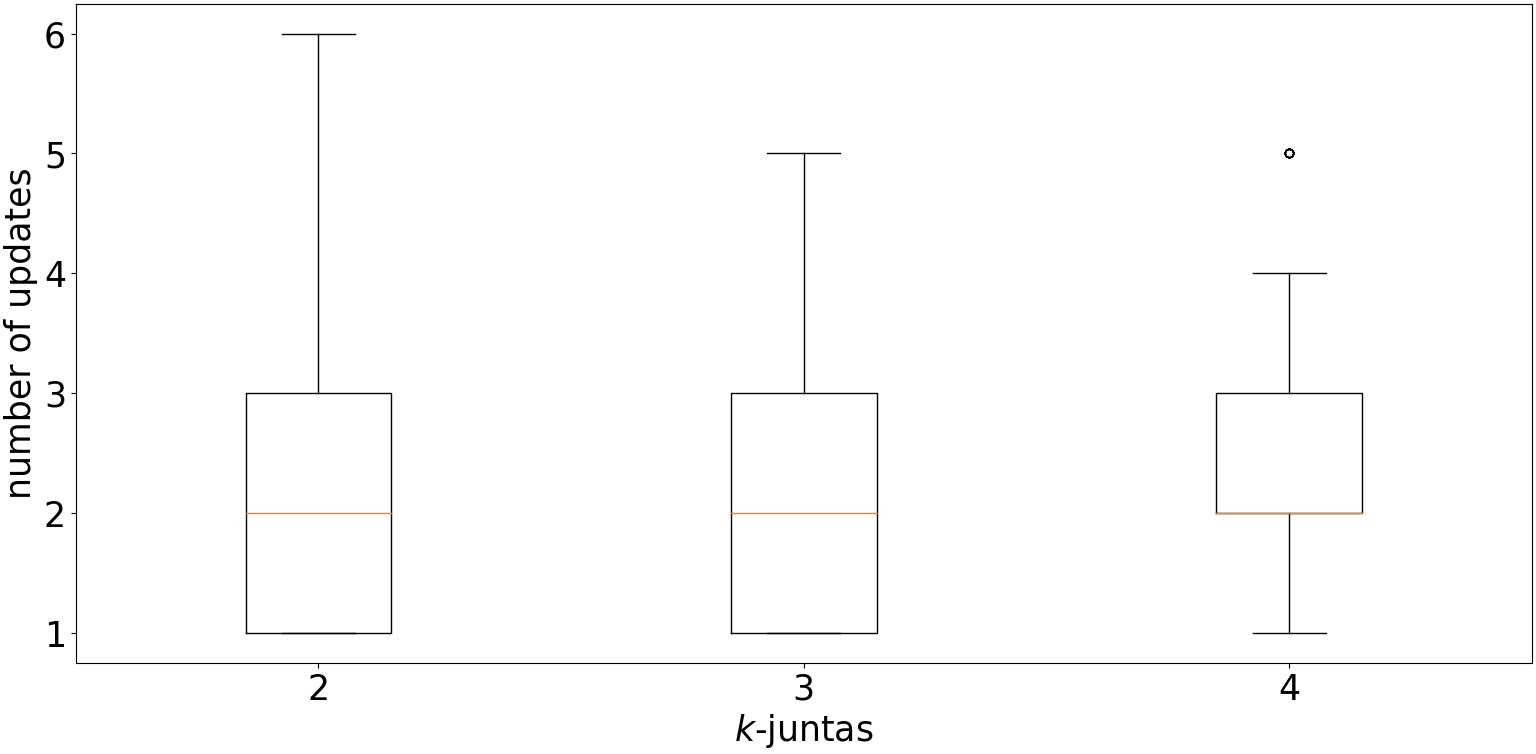}
        \caption{Number of updates to learn $k$-juntas for $n=5$ and $k \in [2 \ldots 4]$}
        \label{fig:compound_updates_juntas_5}
    \end{subfigure}
    \hfill
    \begin{subfigure}{0.49\textwidth}
        \includegraphics[width=\textwidth]{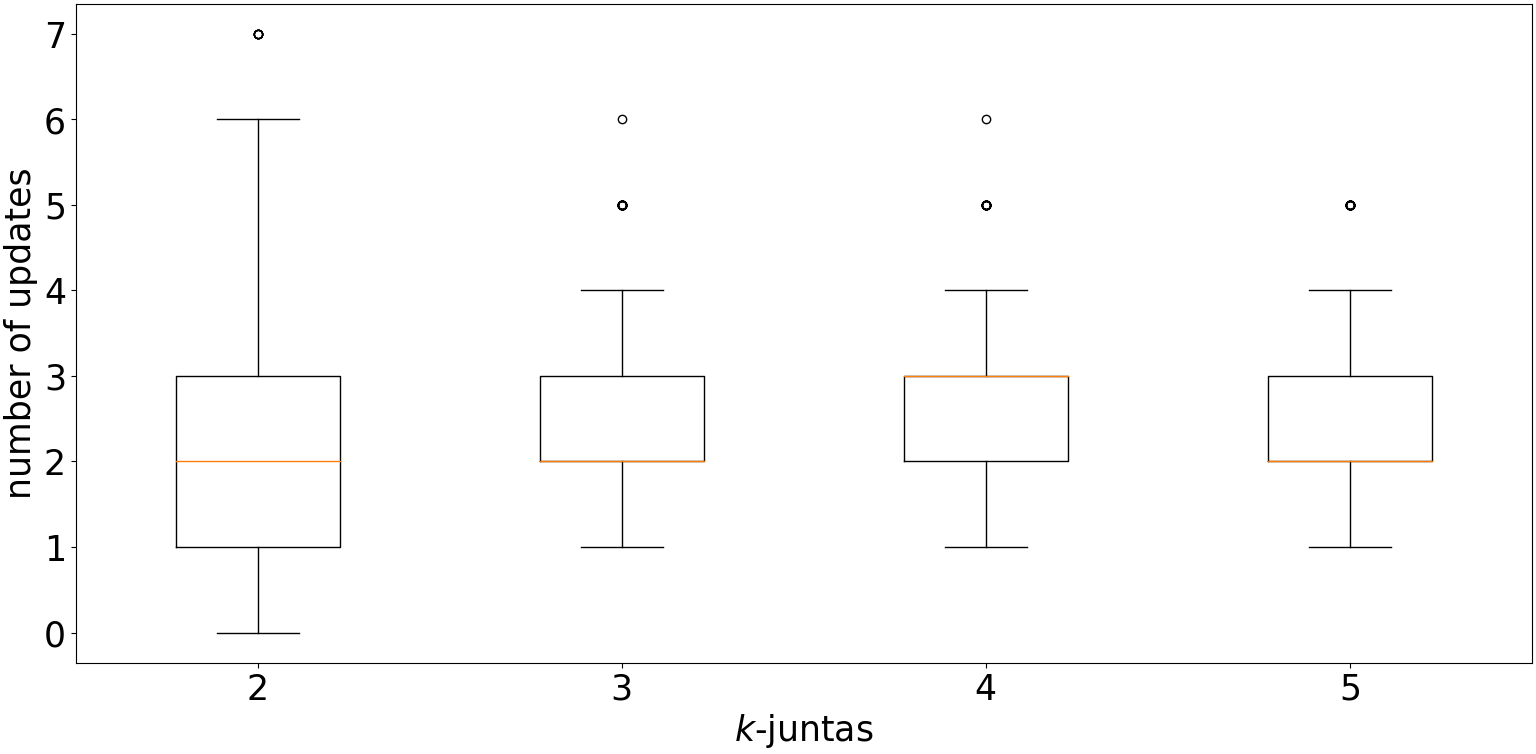}
        \caption{Number of updates to learn $k$-juntas for $n=6$ and $k \in [2 \ldots 5]$}
        \label{fig:compound_updates_juntas_6}
    \end{subfigure}
    \begin{subfigure}{0.49\textwidth}
        \includegraphics[width=\textwidth]{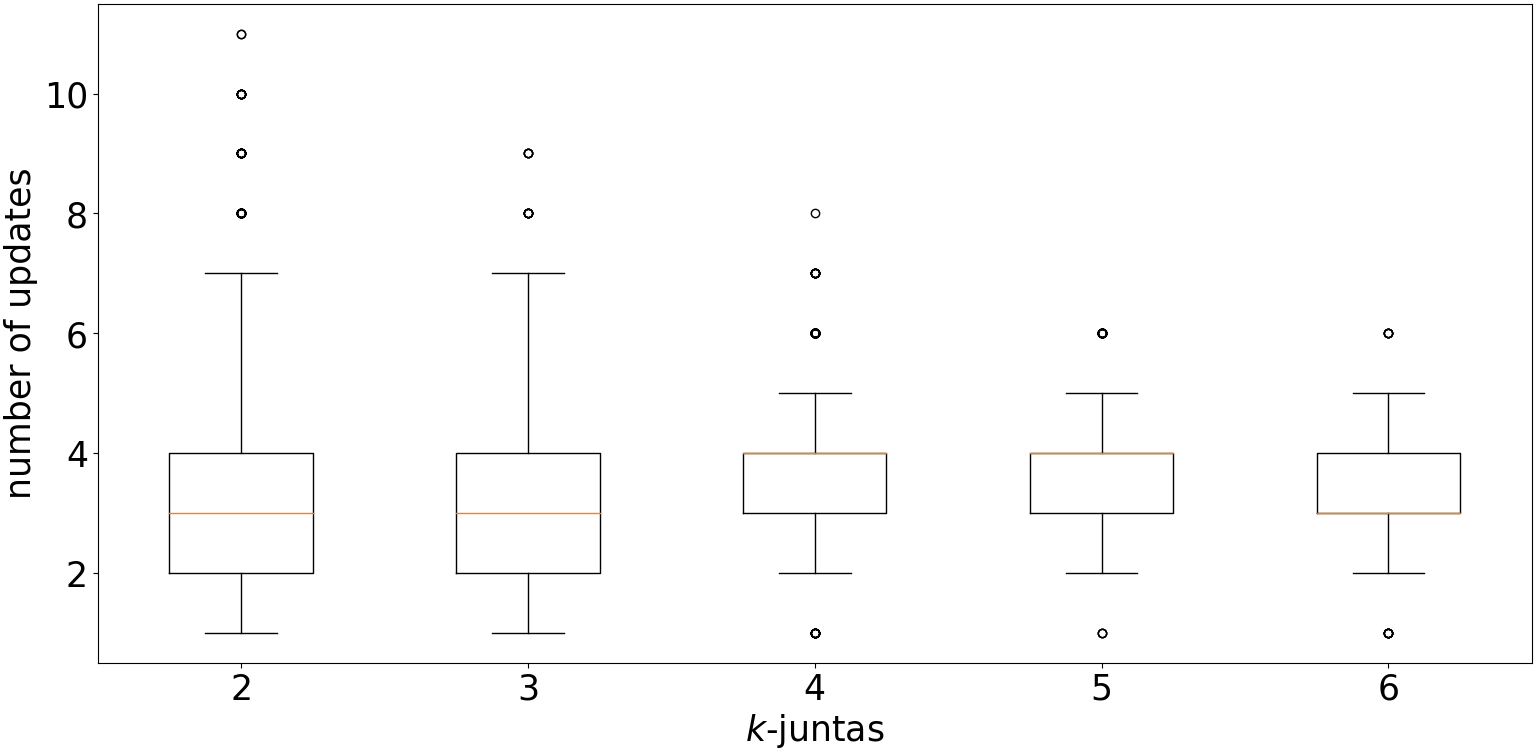}
        \caption{Number of updates to learn $k$-juntas for $n=7$ and $k \in [2 \ldots 6]$}
        \label{fig:compound_updates_juntas_7}
    \end{subfigure}
    \hfill
    \begin{subfigure}{0.49\textwidth}
        \includegraphics[width=\textwidth]{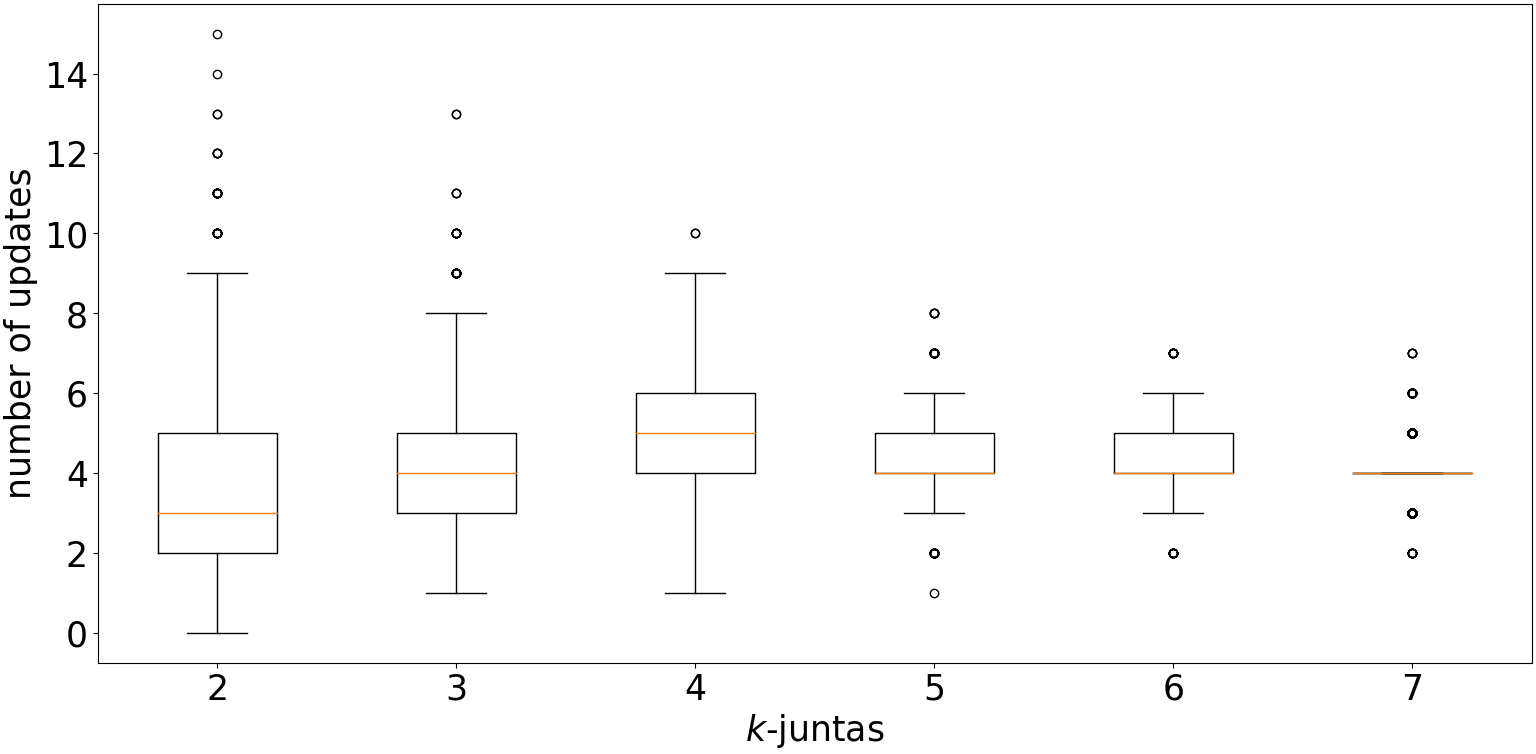}
        \caption{Number of updates to learn $k$-juntas for $n=8$ and $k \in [2 \ldots 7]$}
        \label{fig:compound_updates_juntas_8}
    \end{subfigure}
    \caption{Number of updates for different pairs $(n,k)$}
    \label{fig:updates_juntas}
\end{figure}

To have an overview of this metric, for a given $n$ and a given $k$, we have aggregated all of the training runs for all of the functions. For a given $n$, this set of data has then been plotted against $k$ using box plots. This way we can visualise the median, first and third quartiles, as depicted by the red line, the lower bound, and the upper bound of the box respectively. The maximum and minimum are indicated by the whiskers and the circles represent outliers. All of these are shown in Figure \ref{fig:updates_juntas}.
From these results, we can see that for a given $n$, the number of samples is indeed in the order of $n$. But more interestingly, it slightly decreases as $k$ increases. This can be explained by the fact that during an update phase, the update algorithm has access to more samples, hence the descent to the principals of the target function is quicker. Another reason comes from the fact that for the algorithm to find a principal when $k$ is small, it will potentially have to go "deeper" as the Hamming weight of a principal is at most $k$.

However, our upper bound of $n$ still holds. From these experiments, we have verified that the total sample complexity is in $O(n^2 2^k)$. In \cite{Atici2007}, the task was also to learn $k$-juntas while given access to a uniform quantum example oracle albeit in the QPAC-learning framework. The complexity of their algorithm was then $O\left(\frac{k}{\epsilon}\log(k)\right)$. Assuming that this algorithm can be applied in the exact learning framework by taking $\epsilon = \frac{1}{2^n}$, we end up with a complexity of $O(2^nk\log(k))$. So in the case where $k \ll n$, our algorithm will perform better.

\section{Conclusion}
In this work, we have devised an algorithm to train a tunable quantum neural network in the exact learning framework with access to a uniform quantum example oracle. We refined it by employing it to learn generic Boolean functions and by comparing it to a naive algorithm. We then adapted this algorithm to learn the class of $k$-juntas. Following the implementation of this approach, we found that the query complexity is in $O(n^2 2^k)$. This complexity is lower than what can be found in the literature in the case where $k \ll n$.

\nocite{*}
\bibliographystyle{unsrt}
\bibliography{generic}
\end{document}